\documentclass[paper]{JHEP3}
\usepackage{epsfig}
\usepackage{graphicx}
\usepackage{xspace}
\usepackage{amssymb}
\usepackage{subfigure}
\usepackage{multirow}

\newcommand{\be}{\begin{equation}}
\newcommand{\ee}{\end{equation}}

\newcommand{\bea}{\begin{eqnarray}}
\newcommand{\eea}{\end{eqnarray}}
\newcommand{\beanon}{\begin{eqnarray*}}
\newcommand{\eeanon}{\end{eqnarray*}}
\newcommand{\ba}{\begin{array}}
\newcommand{\ea}{\end{array}}
\newcommand{\bd}{\begin{description}}
\newcommand{\ed}{\end{description}}
\newcommand{\bi}{\begin{itemize}}
\newcommand{\ei}{\end{itemize}}
\newcommand{\ben}{\begin{enumerate}}
\newcommand{\een}{\end{enumerate}}
\newcommand{\bc}{\begin{center}}
\newcommand{\ec}{\end{center}}

\newcommand{\ordEW}{\mathcal{O}(\alpha_{\scriptscriptstyle EM}^6)\xspace}
\newcommand{\ordQCD}{\mathcal{O}(\alpha_{\scriptscriptstyle EM}^4
  \alpha_{\scriptscriptstyle S}^2)\xspace}
\newcommand{\ordQCDsq}{\mathcal{O}(\alpha_{\scriptscriptstyle EM}^2
  \alpha_{\scriptscriptstyle S}^4)\xspace}

\newcommand{\eqn}[1]{Eq.(\ref{#1})}
\newcommand{\eqns}[2]{Eqs.(\ref{#1}--\ref{#2})}

\newcommand{\tbn}[1]{Tab.~\ref{#1}}

\newcommand{\fig}[1]{Fig.~\ref{#1}}
\newcommand{\figs}[2]{Figs.~\ref{#1}--\ref{#2}}
\newcommand{\figsc}[2]{Figs.~\ref{#1},~\ref{#2}}
\newcommand{\sect}[1]{Sect.~\ref{#1}}

\newcommand{\Phantom}{{\tt PHANTOM}\xspace}

\newcommand{\MadEvent}{{\tt MADEVENT}\xspace}

\def\pl #1 #2 #3 {{\it Phys.~Lett.} {\bf#1} (#2) #3}   
\def\np #1 #2 #3 {{\it Nucl.~Phys.} {\bf#1} (#2) #3}
\def\zp #1 #2 #3 {{\it Z.~Phys.} {\bf#1} (#2) #3}
\def\pr #1 #2 #3 {{\it Phys.~Rev.} {\bf#1} (#2) #3}
\def\prep #1 #2 #3 {{\it Phys.~Rep.} {\bf#1} (#2) #3}
\def\prl #1 #2 #3 {{\it Phys.~Rev.~Lett.} {\bf#1} (#2) #3}
\def\intj #1 #2 #3 {{\it Int. J. Mod. Phys.} {\bf#1} (#2) #3}
\def\mpl #1 #2 #3 {{\it Mod.~Phys.~Lett.} {\bf#1} (#2) #3}
\def\rmp #1 #2 #3 {{\it Rev. Mod. Phys.} {\bf#1} (#2) #3}
\def\cpc #1 #2 #3 {{\it Comp. Phys. Commun.} {\bf#1} (#2) #3}
\def\epj #1 #2 #3 {{\it Eur. Phys. J.} {\bf#1} (#2) #3}
\def\jhep #1 #2 #3 {{\it JHEP} {\bf#1} (#2) #3}

\title{
How well can the LHC distinguish between the SM 
light Higgs scenario, a composite Higgs and the Higgsless case using
VV scattering channels? 
}

\author{
Alessandro Ballestrero$^a$,
Giuseppe Bevilacqua$^c$
Diogo Buarque Franzosi$^{a,b}$ and
Ezio Maina$^{a,b}$\\
$^a$ INFN, Sezione di Torino, Italy,\\
Via Giuria 1, 10125 Torino, Italy,\\
$^b$ Dipartimento di Fisica Teorica, Universit\`a di Torino, Italy\\
Via Giuria 1, 10125 Torino, Italy,\\
$^c$ Institute of Nuclear Physics, NCSR Demokritos,\\
Patriarchou Gregoriou \& Neapoleos Str., 15310 Agia Paraskevi, Athens,
Greece.\\
Email: ballestrero@to.infn.it, bevilacqua@inp.demokritos.gr, buarque@to.infn.it, 
maina@to.infn.it.
}

\preprint{DFTT 54/2009}

\abstract{
A complete parton level analysis of $\ell^+\ell^-$ + four jets ($\ell = e,\mu$)
and $3\ell\nu$ +
two jets production at the LHC is presented, including all processes at order
$\ordEW$, $\ordQCD$ and $\ordQCDsq$ when appropriate. The infinite Higgs mass
scenario, which is considered as a benchmark for strong scattering theories and
is the limiting case for composite Higgs models, and one example of a model
incorporating a Strongly Interacting Light Higgs are confronted with the
Standard Model light Higgs predictions. This analysis is combined with the
results in the $\ell\nu$ + four jets channel presented in a previous paper, in
order to determine whether a composite Higgs signal can be detected as an excess
of events in boson--boson scattering.}

\begin{document}

\section{Introduction}
\label{sec:intro}

The Standard Model (SM) describes Electroweak Symmetry Breaking (EWSB) in the
most simple and economical fashion through a single complex Higgs doublet, with
only a neutral scalar field in the spectrum. The fit of EW precision data is in
agreement with the SM predictions to an unprecedented accuracy and gives an
upper limit on the Higgs mass of about 182 GeV \cite{lepewwg07}. Direct searches
at LEP2 imply
$m_H > 114 \mbox{ GeV}$ \cite{lepewwg} while, more recently D0 and CDF 
at the Tevatron have excluded at
95\%CL a SM Higgs in the range $160\mbox{ GeV} < m_H <  170 \mbox{ GeV}$
\cite{TevatronHiggsLimit}.
The LHC will have the task to reveal whether this
minimal realization of EWSB takes place in Nature or a more complex structure is
present\footnote{ Detailed reviews and extensive bibliographies can be found in
Refs.~\cite{HiggsLHC,djouadi-rev1,ATLAS-TDR,Houches2003,CMS-TDR}}.

If the Higgs field is not found, scattering processes between longitudinally
polarized vector bosons will play a prominent role because, without a Higgs, the
corresponding amplitudes grow with energy and violate perturbative unitarity at
about one TeV \cite{reviews}, requiring new physics in the energy range accessible to the
LHC.

Many alternative mechanisms of EWSB have been explored. We will not try to
summarize the different models and simply refer to the literature. For our
purpose we will only remark that it is conceivable and widely discussed
\cite{H_Goldstone,LittleH1,LittlestH,gaugeHiggsU1,gaugeHiggsU2,HologHiggs,LittleHcustodial}
that composite states are responsible for EWSB as nicely
recently reviewed in Ref.~\cite{Giudice_rev08}. These theories are characterized by the presence
of new states which could be produced at the LHC, if light enough.

In view of
the large number of different proposals it is useful to determine the model
independent features of this class of theories.
There has been recent progress in this area \cite{Contino07,Giudice:2007fh,Barbieri},
using the effective
theory language \cite{EEWL}. In Ref.~\cite{Giudice:2007fh}
it has been pointed out that, if EWSB is
triggered by a light composite Higgs which is a pseudo--Goldstone boson related
to some large scale strongly interacting dynamics, the growth with energy of the
vector boson scattering amplitudes typical of Higgsless models might not be
completely canceled by Higgs exchange diagrams but only slowed down because of
the modified couplings between the Higgs field and the vector bosons with respect
to the SM ones. Such a Higgs has been called Strongly Interacting Light Higgs
(SILH). If the mass of the Higgs boson becomes larger than the typical energy
scale at which boson--boson scattering is probed, the contribution of the Higgs
exchange diagrams decreases and completely vanishes, in the Unitary gauge we
employ throughout our calculation, in the limit of an infinite mass which we
will refer to as the Higgsless case. As a consequence, the scattering cross
section for an infinitely massive Higgs in the SM represents, at large energies,
an upper limit for VV scattering processes in SILH models, and can be taken as a
benchmark for the observability of signals of strong scattering and of Higgs
compositeness in boson--boson reactions. On the other hand the Higgsless case is
also representative of models in which heavy resonances which
unitarize boson--boson scattering  are present but cannot be directly detected at the
LHC.

Scattering processes among vector bosons have been scrutinized since a long time
\cite{history1,history2}. In Ref.~\cite{Accomando:2005hz,Accomando:2006vj}
an analysis of $\ell\nu$ + four jets and $\ell^+\ell^-$
+ four jets production at the LHC has been presented, with the limitation of
taking into account only purely electroweak processes. Preliminary results
concerning the inclusion of the $\ordQCD$ background, which include $V V + 2j$ and
top--antitop production have appeared in Ref.~\cite{Ambroglini:2009mg}.
A preliminary analysis in the Equivalent Vector Boson Approximation of the observability
of partial unitarization of longitudinal vector boson scattering in SILH models
at the LHC can be found in Ref.~\cite{Cheung:2008zh}.
In the last few years QCD
corrections to boson--boson production via vector boson fusion \cite{JagerOleariZeppenfeld}
at the LHC
have been computed and turn out to be below 10\%. Recently, VBFNLO \cite{Arnold:2008rz}
a Monte
Carlo program for vector boson fusion, double and triple vector boson production
at NLO QCD accuracy, limited to the leptonic decays of vector bosons, has been
released.

In Ref.~\cite{Ballestrero:2008gf}
a complete parton level analysis of $\ell\nu$ + four jets production
at the LHC, including all processes at order $\ordEW$, $\ordQCD$ and $\ordQCDsq$
has been presented, comparing a typical SM light Higgs scenario with the
Higgsless case. It was noted that the $\ordQCDsq$ $W$ + 4j background is so
large to make the usual approach of comparing the number of events in the two
scenarios at large invariant masses rather ineffective. An alternative strategy
was suggested, namely to focus on the invariant mass distribution of the two
central jets which is characterized by a peak corresponding to the decays of
vector bosons which is present in the vector--vector scattering signal and by an
essentially flat background produced by $\ordQCDsq$ $W +4j$ processes. The latter
can be measured from the sidebands drastically decreasing the theoretical
uncertainties. In Ref.~\cite{Ballestrero:2008gf}
the probability of observing a signal of new physics
in the $W + 4j$ channel was estimated constructing the probability distribution
of experimental results both in the SM and in the Higgsless case, taking into
account the statistical uncertainties for both signals and background and the
theoretical uncertainty for the signal alone, assuming that the background can
be extrapolated from the measured sidebands. The probability that the benchmark
Higgsless scenario would result in an experimental outcome outside the SM 95\%
exclusion limit, assuming a 200 fb$^{-1}$ luminosity for both the muon and
electron decay channel, turned out to be about 97\%.

In this paper we complete our study examining at parton level the processes $pp
\rightarrow \ell^+\ell^- + 4j$ and $pp \rightarrow 3\ell\nu + 2j$, including all
irreducible backgrounds contributing to these six parton final states. 
These are the only vector--vector scattering reactions,
together with $pp \rightarrow \ell\nu +4j$ mentioned above, in which 
the presence of at least one pair of leptons in the final state ensures
a sufficiently clean determination of the invariant mass of the outgoing vector
bosons, up to the ambiguity related to the momentum reconstruction of the
unobserved neutrino. 
The latter can be however be estimated by imposing the condition that the
missing momentum and the charged lepton reconstruct the W mass. It is here
implied that the $pp \rightarrow 4\ell+2j$ channel, while allowing a more precise
reconstruction of the $VV$ mass, has a negligible rate at the LHC at large
invariant masses of the four-lepton system due to the small branching ratio
of the leptonic $Z$ decays.
The invariant mass $M_{VV}$ of the two outgoing bosons is the analogue of the 
center--of--mass energy for on--shell vector boson scattering: differences
between the Higgsless scenario and the SM predictions increase, for the processes
we analyze, with the boson
pair invariant mass.

In this paper we consider three scenarios: a light Higgs SM framework with $M_H = 200$ GeV,
one instance of the SILH models which we will describe shortly and an infinite
mass Higgs scenario. We combine the results for $pp \rightarrow \ell^+\ell^- + 4j$
and $pp \rightarrow 3\ell\nu + 2j$ with those obtained in Ref.~\cite{Ballestrero:2008gf}
for the $pp \rightarrow \ell\nu +4j$ channel in order to
obtain a global estimate, including all three channels, of the probability of
distinguishing the considered
Beyond Standard Model (BSM) cases from the SM scenario.

\section{The Strongly Interacting Light Higgs scenario}
\label{sec:SILH}

The effective field theory approach \cite{EEWL} is a powerful tool for describing the
low energy dynamics of systems with broken symmetries. It provides a systematic
expansion of the full unknown Lagrangian in terms of the fields which are
relevant at scales much lower than the symmetry breaking scale. In Ref.~\cite{Giudice:2007fh}
this general framework has been employed to describe the main features of the
models of EWSB in which the Higgs field can be identified with a
pseudo--Goldstone boson of some broken strong interaction at high energy. Models
which fall into this class are for instance the Holographic Higgs \cite{HologHiggs}, the
Little Higgs of Ref.~\cite{LittleHcustodial} and the Littlest Higgs \cite{LittlestH}.

These theories generally predict new resonances and their search may well be
the best way to detect New Physics effects at the LHC. However they also lead to
modifications of boson--boson scattering which might provide further evidence and
possibly be the only accessible effects if the new resonances are heavy.

The leading low energy effects are described by two parameters (one responsible
for a
universal modification of all Higgs couplings, and the other one for a universal
modification of Higgs couplings to fermions) characterized by the ratio $v^2/f^2
= \xi$, where $v$ is the Higgs vacuum expectation value and $f$ is the
$\sigma$--model scale.
The natural range of the $\xi$ parameter is between $\xi=0$ and $\xi=1$ which correspond
respectively to the limiting cases of the Standard Model and of technicolor theories.
Because of the modified Higgs couplings, longitudinal
gauge--boson scattering amplitudes violate unitarity at high energy, even in the
presence of a light Higgs \cite{Giudice:2007fh}. The $E^2$--growing amplitude is
typically smaller than in the Higgsless case and the violation is moved to a
larger energy regime.

Therefore, even if a light Higgs is discovered, boson--boson scattering is a
crucial process to study and can give us useful information on the nature of
the Higgs boson. It is worth pointing out that, in this framework, since the
Higgs can be viewed as an approximate fourth Goldstone boson, its properties are
related to those of the exact (eaten) Goldstone bosons. Strong gauge--boson
scattering will be accompanied by strong Higgs pair production \cite{Giudice:2007fh}.

The effective Lagrangian approach of Ref.~\cite{Giudice:2007fh} is valid for small
values of $\xi$,  while larger values demand a more detailed description of the
particular model at hand. Such a Lagrangian leads to a modification of the Higgs 
couplings by a factor $1/\sqrt{1+c_H\xi}$, which can be reabsorbed in a Higgs
propagator modification by a factor $1/(1+c_H\xi)$ in boson boson scattering
studies. $c_H$ is a pure number of order unity 
\cite{LittlestH,HologHiggs, LittleHcustodial,Giudice:2007fh}.
For the present study we have selected the value $c_H\xi = 1$ which we intend
as a possible upper limit for the model independent lagrangian description of
Ref.~\cite{Giudice:2007fh}.

\section{Outline of the analysis}
\label{sec:Outline_of_the_analysis}

The observation of strong boson--boson scattering as an excess of events compared
to the SM prediction requires, as an essential condition, that a signal of VV
scattering is extracted from the background. At the same time the selection
strategy must be capable to maximize the differences between the light Higgs and
the BSM cases.

\begin{figure}
\begin{center}
\mbox{\epsfig{file=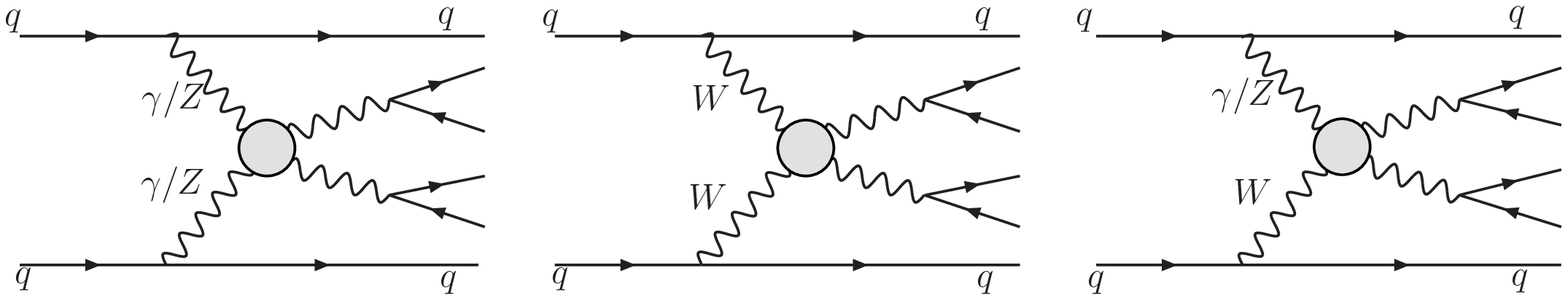,width=12.cm}}
\caption{ Vector boson fusion processes.}
\label{VV-diag}
\end{center}
\end{figure}

\begin{figure}
\begin{center}
\mbox{\epsfig{file=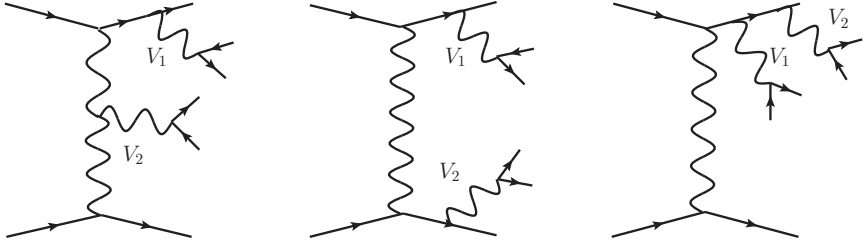,width=12.cm}}
\caption{ Two vector boson production without fusion.}
\label{nonreso-diag}
\end{center}
\end{figure}

\begin{figure}
\begin{center}
\mbox{\epsfig{file=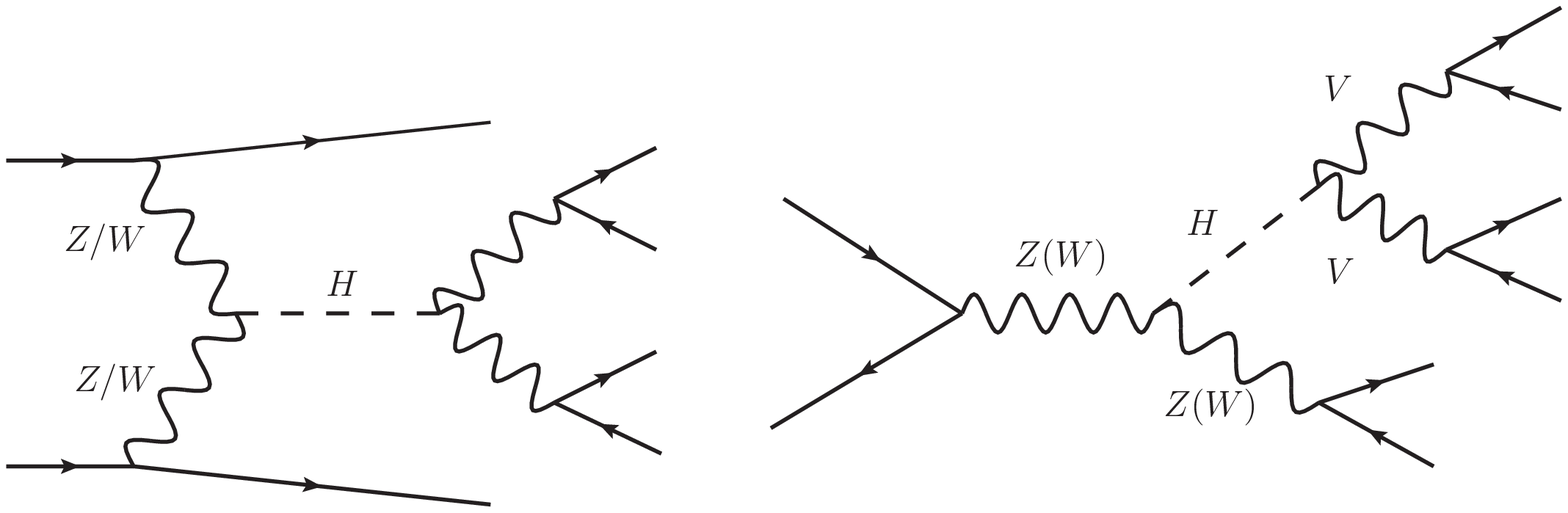,width=12.cm}}
\caption{Higgs boson production via vector boson fusion
  and Higgsstrahlung.}
\label{higgs-diag}
\end{center}
\end{figure}

\begin{figure}
\begin{center}
\mbox{\epsfig{file=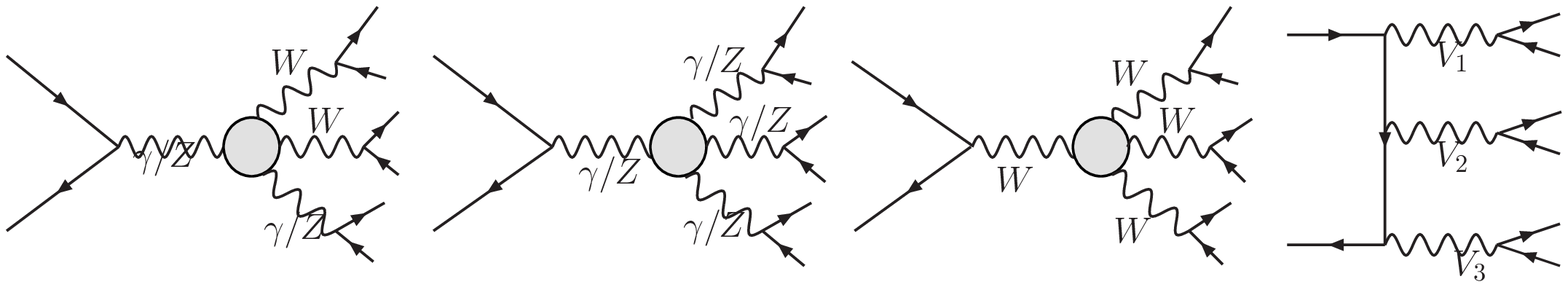,width=14.cm}}
\caption{ Three vector boson production.}
\label{tgc-diag}
\end{center}
\end{figure}

Three(two) perturbative orders contribute to the background
for $\ell^+ \ell^- + 4j$($3\ell\nu + 2j$) final states. At $\ordEW$ there are a
large number of diagrams which cannot be interpreted as boson--boson scattering
and which cannot be separated in any sensible way from the scattering type
diagrams due to large cancellations between the two sets \cite{Accomando:2006vj}.
For illustrative purpose a number of diagrams are shown in
\figs{VV-diag}{fig:diag_as4}. \fig{VV-diag} illustrates the class of diagrams which include
boson--boson scattering subdiagrams while \fig{nonreso-diag} presents some representative
$\ordEW$ diagrams in which two vector bosons are produced but no fusion takes place.
\fig{higgs-diag} and \fig{tgc-diag} show diagrams which describe Higgs production and
three vector boson production.

\begin{figure}[h!tb]
\centering
\hspace*{-0.5cm}
\epsfig{file=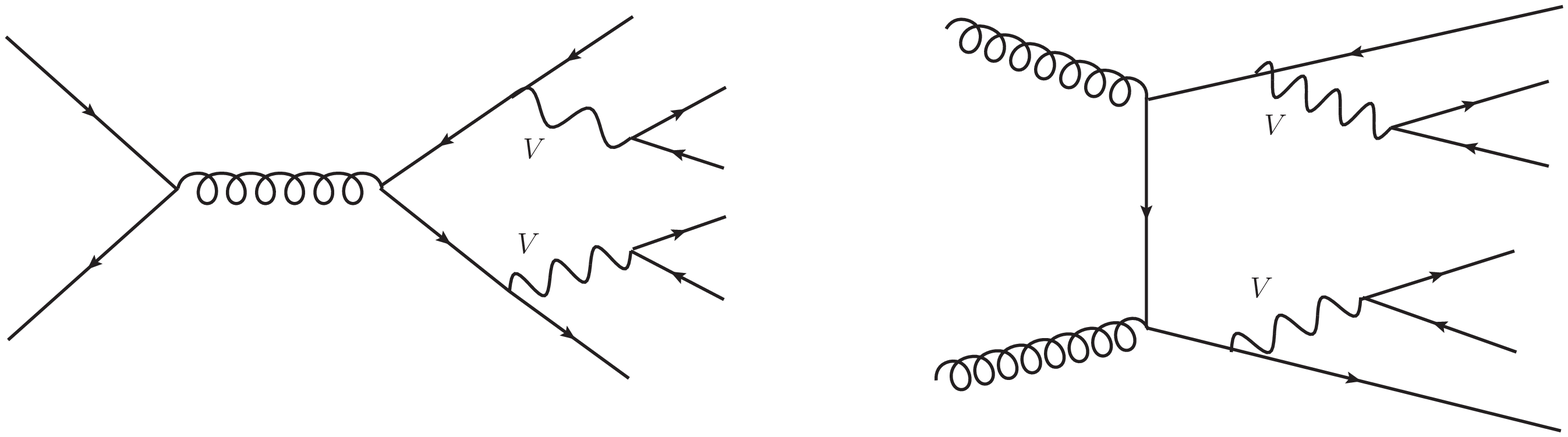,width=7.5cm}
\hspace*{0.2cm}
\epsfig{file=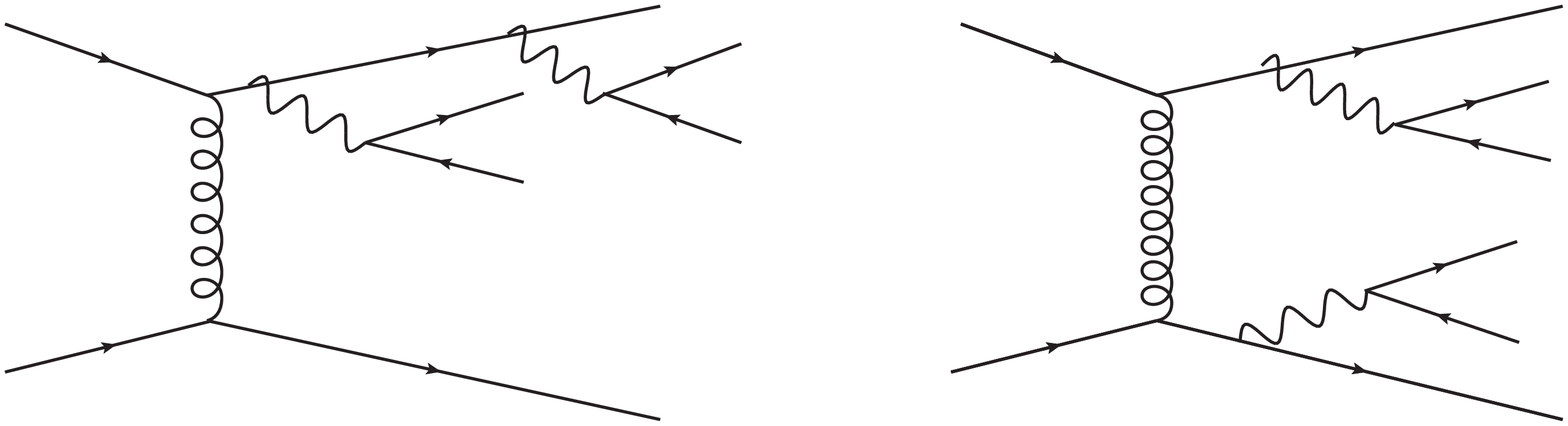,width=7.5cm}
\hspace*{-0.7cm}
\begin{picture}(0,0) (0,0)
  \put(-394,-5) {\small{(a)}}
  \put(-274,-5) {\small{(b)}}
  \put(-159,-5) {\small{(c)}}
  \put(-43,-5) {\small{(d)}}
\end{picture}
\caption{Examples of contributions to the QCD irreducible background: $VV+2j$}
\label{fig:VBSbckgr_QCD}
\end{figure}

\begin{figure}[htb]
\centering
\mbox{
\includegraphics*[width=0.29\textwidth]{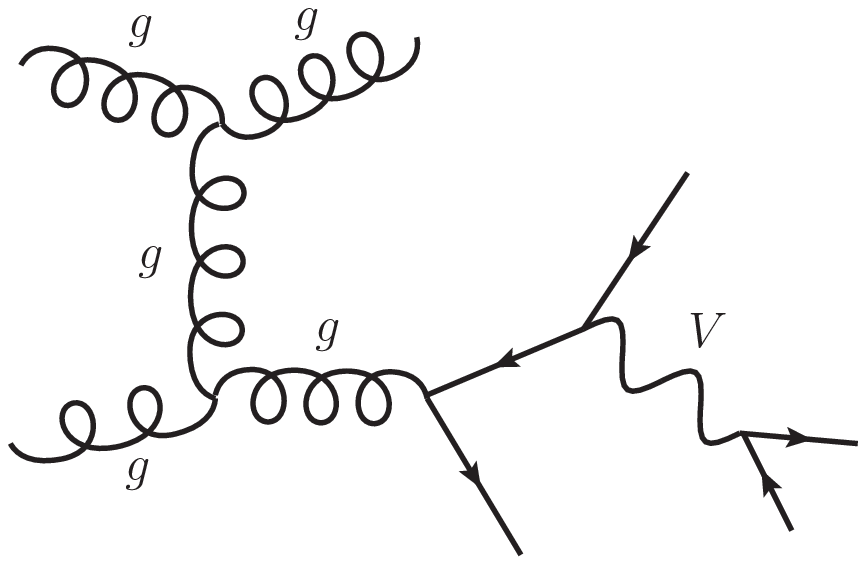}\hspace*{0.5cm}
\includegraphics*[width=0.20\textwidth]{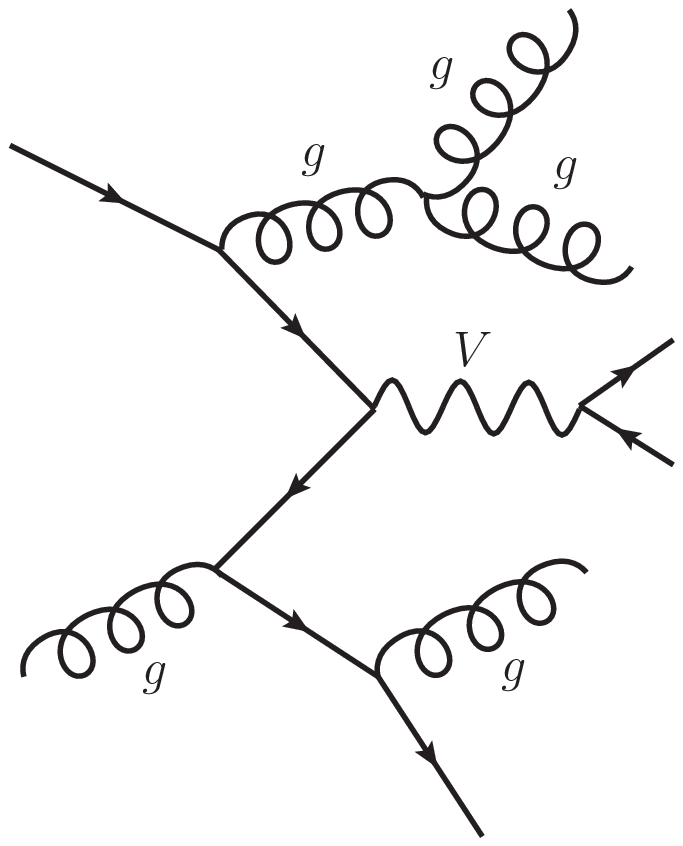}\hspace*{0.6cm}
\includegraphics*[width=0.18\textwidth]{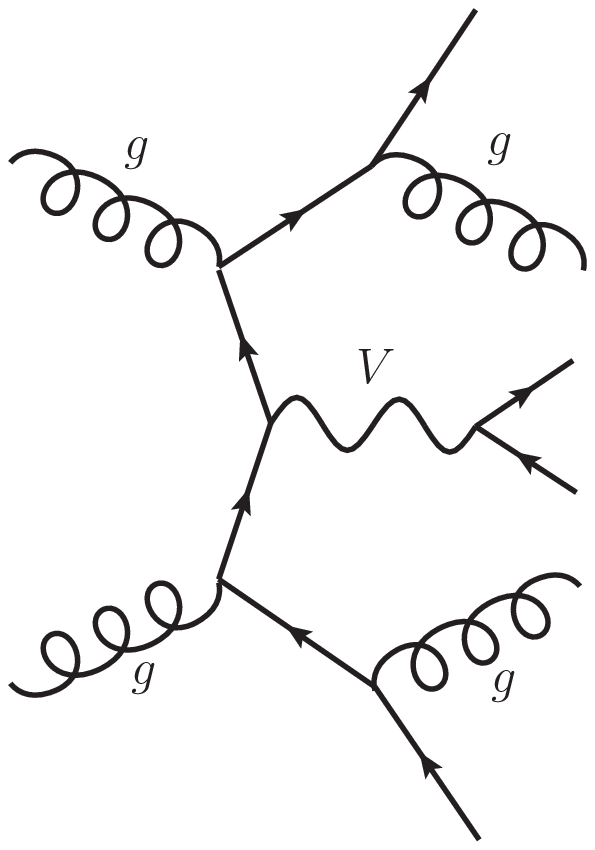}\hspace*{0.6cm}
\includegraphics*[width=0.19\textwidth]{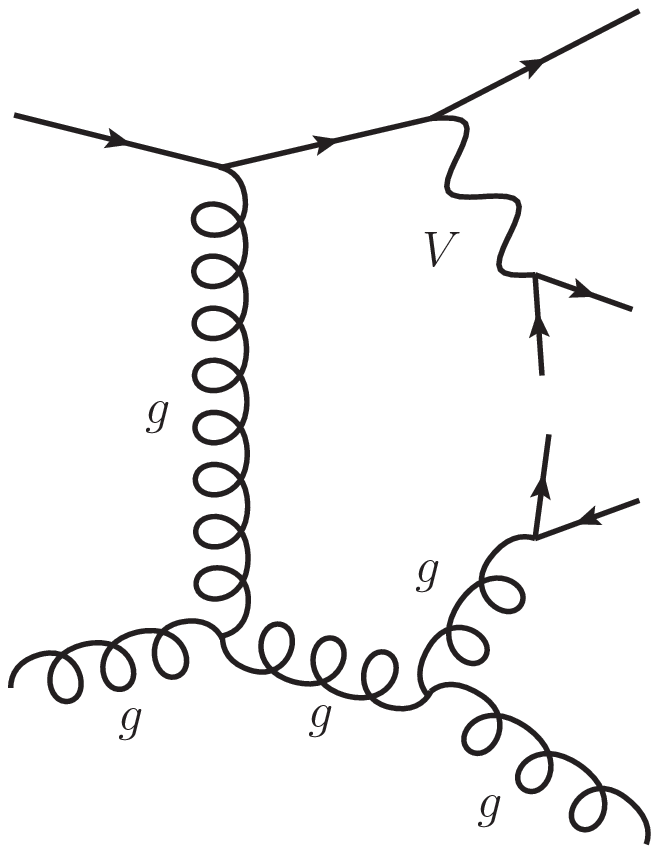}
}\\[0.2cm]
\begin{picture}(0,0) (0,0)
  \put(-180,-5) {\small{(a)}}
  \put(-35,-5) {\small{(b)}}
  \put(65,-5) {\small{(c)}}
  \put(160,-5) {\small{(d)}}
\end{picture}
\caption{}
Representative Feynman diagrams for the $\ordQCDsq$, $V+4j$
production processes at the LHC.
\label{fig:diag_as4}
\end{figure}

At $\ordQCD$ we
have to deal with the production of two electroweak bosons plus two jets without
any scattering contribution.
Some representative diagrams are shown in \fig{fig:VBSbckgr_QCD} which also
includes diagrams describing $t\overline{t}$ production which contribute only
to the $\ell \nu + 4j$ channel.
For the
$\ell^+ \ell^- + 4j$ final state there are contributions at $\ordQCDsq$ in which
only one electroweak boson is effectively produced, while the additional jets,
which, taken in pairs, do not peak at any particular mass, populate the full
available phase space with a production rate which is much larger than the
signal one.
A few diagrams in this set are presented in \fig{fig:diag_as4}.

The first step is concerned with the identification of a suitable kinematic
signature which allows to capture the essence of $VV$ scattering. The selection
of events widely separated in pseudorapidity is a well established technique for
enhancing the scattering contributions at the LHC \cite{history1,history2}.
Looking at the topology of the diagrams embedding the gauge boson scattering as
a subprocess, one concludes that it is appropriate to associate the two most
forward/backward jets to the tag quarks which radiate the bosons which initiate
$VV$ scattering. As shown in Ref.~\cite{Ballestrero:2008gf} a powerful tool to
increase the separation between the SM predictions and those of the Higgsless
scenario is provided, at large invariant masses,
by the request that the vector bosons and their decay
products are in the central part of the detector since the vector bosons in the
Higgsless case have smaller rapidities and larger momenta than in the presence
of a light Higgs. The main purpose of this kinematic selection is to isolate a
sample of genuine $VV + 2j$ events while suppressing the contribution of
irreducible backgrounds such as three boson production or top quark production.

Having isolated a sample of candidate scattering events, one needs to define an
observable quantity which is as susceptible as possible to the details of the
mechanism of EWSB in order to maximize the sensitivity to effects of alternative
models such as strong scattering. For the $\ell^+ \ell^- + 4j$ channel this task
is straightforward only apparently. As already mentioned,
the QCD background is expected to provide a significant number of fake $VV+2j$ events
as a consequence of the large cross
section. The classical approach is to focus on the invariant mass distribution
of the final state boson pair.
The huge QCD background, with its large scale uncertainty,
makes this procedure rather dubious for the $\ell^+\ell^- + 4j$ case. 
Detecting signals of EWSB in the vector pair mass distribution remains
a formidable challenge. The analysis of Ref.~\cite{Ballestrero:2008gf} shows that,
even focusing the attention on the high invariant vector pair mass region and
requiring the mass of the two central jets to be close to the mass of a vector boson,
the expected signal over background ratio is of the order of 1/10.
This way of measuring the signal, typical of a counting experiment,
finds a substantial obstacle in the spread of the interesting excess of events
over a wide region in $M_{VV}$.

A possible way out for this problem is to look instead at the invariant mass of
the two central jets ($M_{j_c j_c}$) for events with large vector pair mass.
Provided a convenient set of kinematic cuts has been applied, the $\ordEW$ +
$\ordQCD$ cross section is dominated by the peaks corresponding to $W$ and $Z$
decays to quarks, while the $\ordQCDsq$ $Z + 4j$ contribution is non--resonant
in this respect. When restricting to the window between 70 and 100 GeV, which
covers completely the $W$ and $Z$ resonances, we find that the $Z + 4j$
distribution is essentially flat and therefore can reliably be measured from the
sidebands of the physical region of interest. This procedure has several advantages.
On one side, it eliminates the theoretical uncertainty associated with
the scale dependence of the $\ordQCDsq$ contribution.
On the other side, it allows to subtract the dominant contribution
to the irreducible QCD background, enhancing the visibility of genuine EWSB
effects. The signal has a very clear signature, a bump in the two central jet mass
distribution, which is much easier to hunt for experimentally than a diffuse
excess in production cross section.

Once the non--resonant background has been subtracted, one is left with a peak
whose size is strictly related to the regime of the EWSB dynamics: a
strongly--coupled scenario would result in a more prominent peak than a
weakly--coupled one. This feature suggests to take the integral of the peak as
the discriminator among different models. At a given collider luminosity, the
number of expected events can be derived. With a slight abuse of language, we
call this number the VBS signal. It is by analyzing the probability density
function (p.d.f.) associated with this discriminator that we can determine, in
the last step, the confidence level for a given experimental result to be or not
to be SM--like, in the same spirit of the statistical procedure adopted for the
search of the Standard Model Higgs boson at LEP \cite{Barate:2003sz}.

For the $3\ell\nu + 2j$ channel the physical picture is straightforward, the
invariant mass of the vector boson pair can be directly measured from the
momenta of the charged leptons and the reconstructed neutrino momentum. Since the
background is much smaller than in the $\ell^+ \ell^- + 4j$ case we use the total
expected number of events as discriminator and estimate the probability that
the experimental results in the BSM models are incompatible with the SM.

\section{Calculation}
\label{sec:calculation}

As discussed in \sect{sec:Outline_of_the_analysis},
three perturbative orders contribute to $\ell^+ \ell^-
+ 4j$ at the LHC, while only two perturbative orders contribute to $3\ell\nu +
2j$. The $\ordEW$ and $\ordQCD$ samples have been generated with \Phantom , a dedicated
tree level Monte Carlo generator
which is documented in Ref.~\cite{Ballestrero:2007xq} while additional material
can be found in Refs.~\cite{ref:Phase,method,phact}.
The $\ordQCDsq$ sample has been produced with \MadEvent \cite{MadeventPaper}.
Both programs generate events in the Les Houches Accord File Format \cite{LHAFF}. In
all samples full $2 \rightarrow 6$ matrix elements, without any production times
decay approximation, have been used. For the LHC we have assumed the design energy
of 14 TeV.
For each perturbative order we have generated a sample of five hundred thousand
unweighted events.

For the Standard Model parameters we use the input values:
\begin{equation}
\begin{array}[b]{lcllcllcl}
\label{eq:SMpar}
\mathrm {M_W} & = & 80.40 , \qquad &
\mathrm {M_Z} & = & 91.187~\mathrm{GeV}, \\
G_{\mu} & = & 1.16639~10^{-5}~\mathrm{GeV}^{-2}, \qquad &
\alpha_s(\mathrm {M_Z})  & = & 0.118\\
\mathrm {M_t} & = & 175.0~\mathrm{GeV}, \qquad &
\mathrm {M_b} & = & 4.8~\mathrm{GeV}. 
\end{array}
\end{equation}

The masses of all other partons have been set to zero. We adopt the standard
$G_{\mu}$--scheme to compute the remaining parameters.

All samples have been generated using CTEQ5L \cite{CTEQ5} parton distribution
functions. For the $\ordEW$ and $\ordQCD$ samples, generated with \Phantom, the
QCD scale has been taken as:

\be
\label{eq:LargeScale}
Q^2 = M_W^2 + \frac{1}{6}\,\sum_{i=1}^6 p_{Ti}^2
\label{scale}
\ee

while for the $\ordQCDsq$ sample the scale has been set to $Q^2 = M_Z^2$. This
difference in the scales conservatively leads to a definite relative enhancement
of the $Z\, + \, 4j$ background. Tests in comparable reactions at $\ordQCDsq$
have shown an increase of about a factor of 1.5 for the processes computed at
$Q^2 = M_Z^2$ with respect to the same processes computed with the larger scale
\eqn{eq:LargeScale}.

We work at parton level with no showering and hadronization. The two jets with
the largest and smallest rapidity are identified as forward and backward tag jet
$(j_f, j_b)$
respectively. The two intermediate jets called central $(j_c)$ and
are considered as candidate vector boson
decay products.

The neutrino momentum is reconstructed according to the usual prescription,
requiring the invariant mass of the $\ell \nu$ pair to be equal to the $W$ boson
nominal mass,

\begin{equation}
\label{eq:nu_reco_equation}
(p^{\ell}+p^{\nu})^2 = M_W^2 ,
\end{equation}

in order to determine the longitudinal component of the neutrino momentum.

For very large Higgs masses, all Born diagrams with Higgs propagators become
completely negligible in the Unitary Gauge we work in. Therefore the Higgsless
model results for all processes coincide with
those in the $M_H \rightarrow \infty$ limit.

The cuts in \tbn{tab:cuts_0} have been applied either at
generation level or as a preliminary step to any further analysis.
They require containment within the active region of the
detectors and minimum transverse momentum for all observed partons; a minimum
mass separation is imposed for all same--family opposite--sign charged leptons,
while a minimum separation both in mass and in $\Delta R$ is required for jet
pairs. At large transverse momentum, jet pairs with mass comparable to the mass
of electroweak bosons or even larger can merge into one single jet when an
angular measure like $\Delta R(jj)$ is adopted for reconstructing jets.
Therefore we have imposed that all partons satisfy $\Delta R(jj) > 0.3$, a value
smaller than usually employed in LHC analyses. However, in \sect{sec:combining}
we discuss in
more detail the effect of removing the angular separation constraint or, on the
contrary, of imposing a more stringent requirement $\Delta R(jj) > 0.5$.
Furthermore, the most forward and most backward jets are required to be
separated by at least four units in rapidity and their combined mass is forced
to be outside the electroweak vector boson mass window; on the contrary the mass
of the two remaining jets, which we will call central in the following, is
required to be compatible with the mass of the weak bosons;
no $jjj$($j\ell\nu$) triplet is
allowed in the neighborhood of the top mass for the
$\ell^+ \ell^- + 4j$($3\ell\nu + 2j$) channel.

\begin{table}[htb]
\begin{center}
\begin{tabular}{|c|}
\hline
\textbf{Acceptance cuts} \\
\hline
\hspace*{1cm} $p_T(\ell^\pm) > 20 \mbox{ GeV}$ \hspace*{1cm} \\
\hline
$|\eta(\ell^\pm)| < 3.0$ \\
\hline
$p_T(j) > 30 \mbox{ GeV}$ \\
\hline
$|\eta(j)| < 6.5$ \\
\hline
$M(jj) > 60 \mbox{ GeV}$ \\
\hline
$M(\ell^+\ell^-) > 20 \mbox{ GeV}$ \\
\hline
$M(j_fj_b)<70 \mbox{ GeV} ; M(j_fj_b)>100 \mbox{ GeV}$ \\
\hline
$|\Delta \eta (j_fj_b)| > 4.0$ \\
\hline
$|M(jjj) - M_{top}| > 15 \mbox{ GeV}$ ($\ell^+ \ell^- + 4j$)\\
$|M(j\ell\nu) - M_{top}| > 15 \mbox{ GeV}$ ($3\ell\nu + 2j$)\\
\hline
\hline
$70 \mbox{ GeV} < M(j_cj_c) < 100 \mbox{ GeV}$ \\
\hline
$\Delta R(jj) > 0.3$ \\
\hline
\end{tabular}
\caption{Standard acceptance cuts applied in the event generation and present in
all results. Here $j = d,u,s,c,b,g$. The last two cuts refer to the $\ell^+
\ell^- + 4j$ and $\ell \nu + 4j$ cases only.} 
\label{tab:cuts_0}
\end{center}
\end{table}

\section{The $\ell^+ \ell^- + 4j$ channel}
\label{sec:Z+4j}

The total cross section for the $\ell^+ \ell^- + 4j$ channel with the generation
cuts in \tbn{tab:cuts_0} is presented in \tbn{tab:res_Z4j_0} as a function of the minimum invariant
mass $M_{cut}$ for the $j_c j_c \ell^+ \ell^-$ system. These results refer to
the mass window $70 \mbox{ GeV} < M(j_cj_c) < 100 \mbox{ GeV}$ and include
all three perturbative orders. In parentheses
the results for the sum of the $\ordEW$ and $\ordQCD$ processes.
\tbn{tab:res_Z4j_0} shows
that the cross section is dominated by the $\ordQCDsq$ contribution. If we
assume a luminosity of $L = 200 \mbox{ fb}^{-1}$ and sum over the electron and
muon channels the difference between the number of events expected for an
infinite mass Higgs and a light one is smaller than the expected statistical
uncertainty for the $\ordQCDsq$ processes and no meaningful separation between
the two cases can be obtained.

\begin{table}[htb]
\begin{tabular}{|p{0.08\textwidth}|p{0.27\textwidth}|p{0.27\textwidth}|p{0.26\textwidth}|}
\hline
$M_{cut}$ 	& \centering no Higgs 	& \centering SILH  & 
					\parbox[t]{0.26\textwidth}{\centering $M_H=200$ GeV}   \\ 
		\cline{2-4}
 (GeV) 		& \centering $\sigma$(fb) & \centering $\sigma$(fb) & \parbox[t]{0.26\textwidth}{\centering $\sigma$(fb)} \\
\hline
600		& 45.8(1.2)	& 45.7(1.06)   & 45.6(1.01)	\\
800		& 23.3(0.605)	& 23.2(0.515)   & 23.2(0.471)	\\
1000		& 12.7(0.318)	& 12.6(0.263)   & 12.6(0.24)	\\
1200		& 7.15(0.177)	& 7.1(0.147)   & 7.1(0.134)	\\
\hline
\end{tabular}
\caption{
Total cross section for the $\mu^+ \mu^- + 4j$ channel after acceptance cuts, \tbn{tab:cuts_0}, in the mass 
window $70 \mbox{ GeV} < M(j_cj_c) < 100 \mbox{ GeV}$.
In parentheses the results for the $\ordEW + \ordQCD$ 
samples.
}
\label{tab:res_Z4j_0}
\end{table}

\begin{figure}[htb]
\centering
\subfigure{
\hspace*{-2.1cm}
\includegraphics*[width=8.3cm,height=6.2cm]{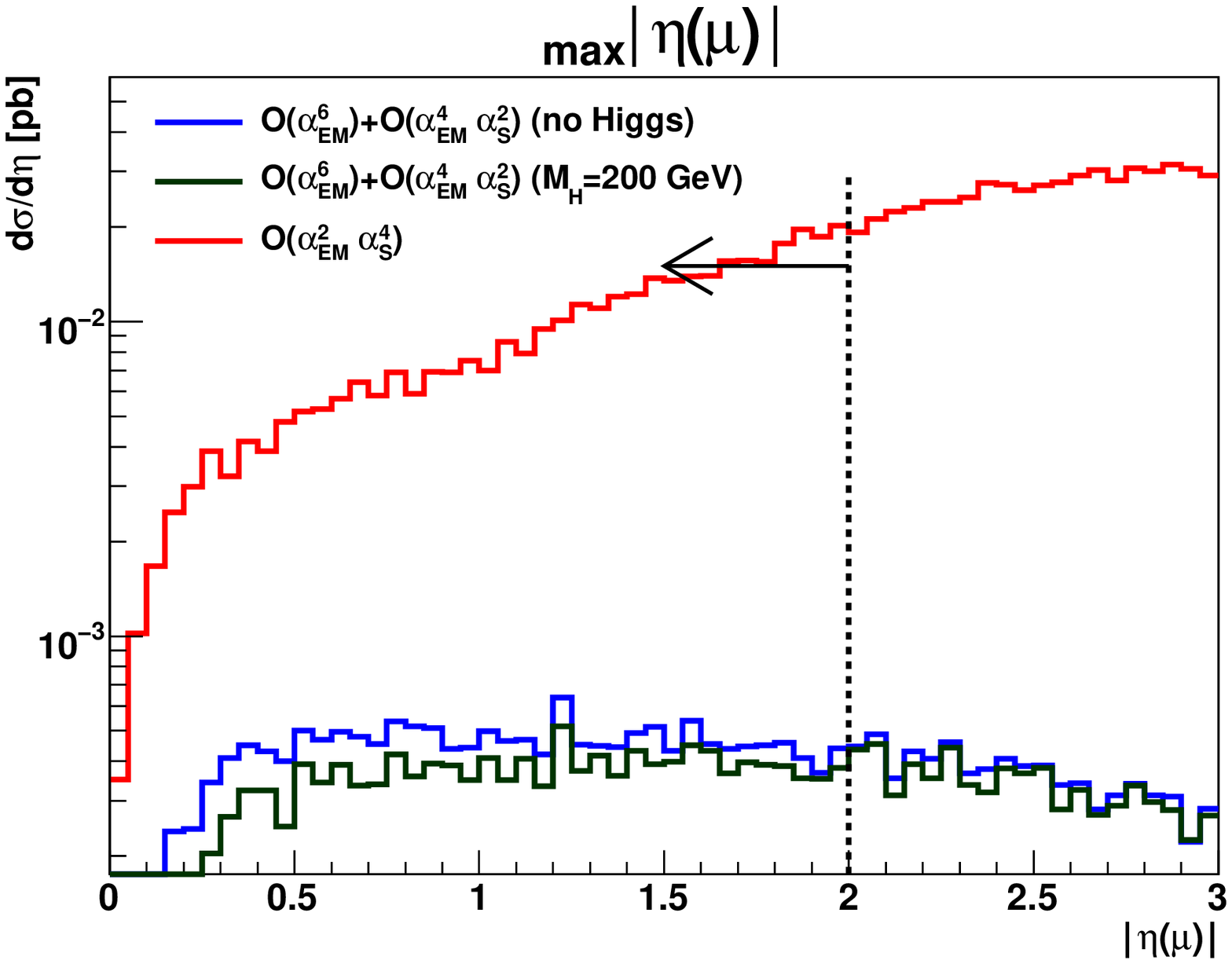}
\hspace*{-0.7cm}
\includegraphics*[width=8.3cm,height=6.2cm]{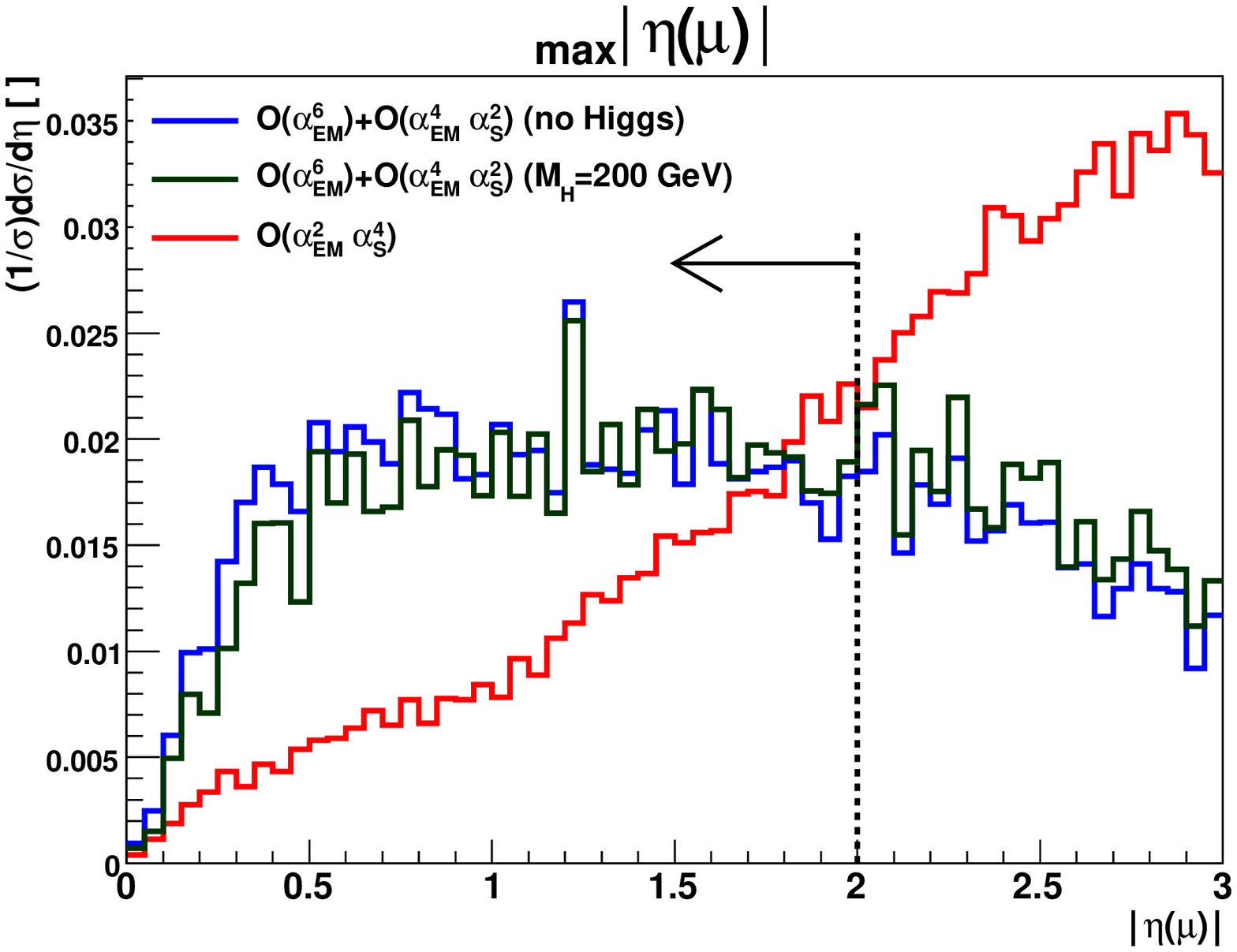}
\hspace*{-3cm}
}
\vspace{-0.4cm}
\subfigure{
\hspace*{-2.1cm}
\includegraphics*[width=8.3cm,height=6.2cm]{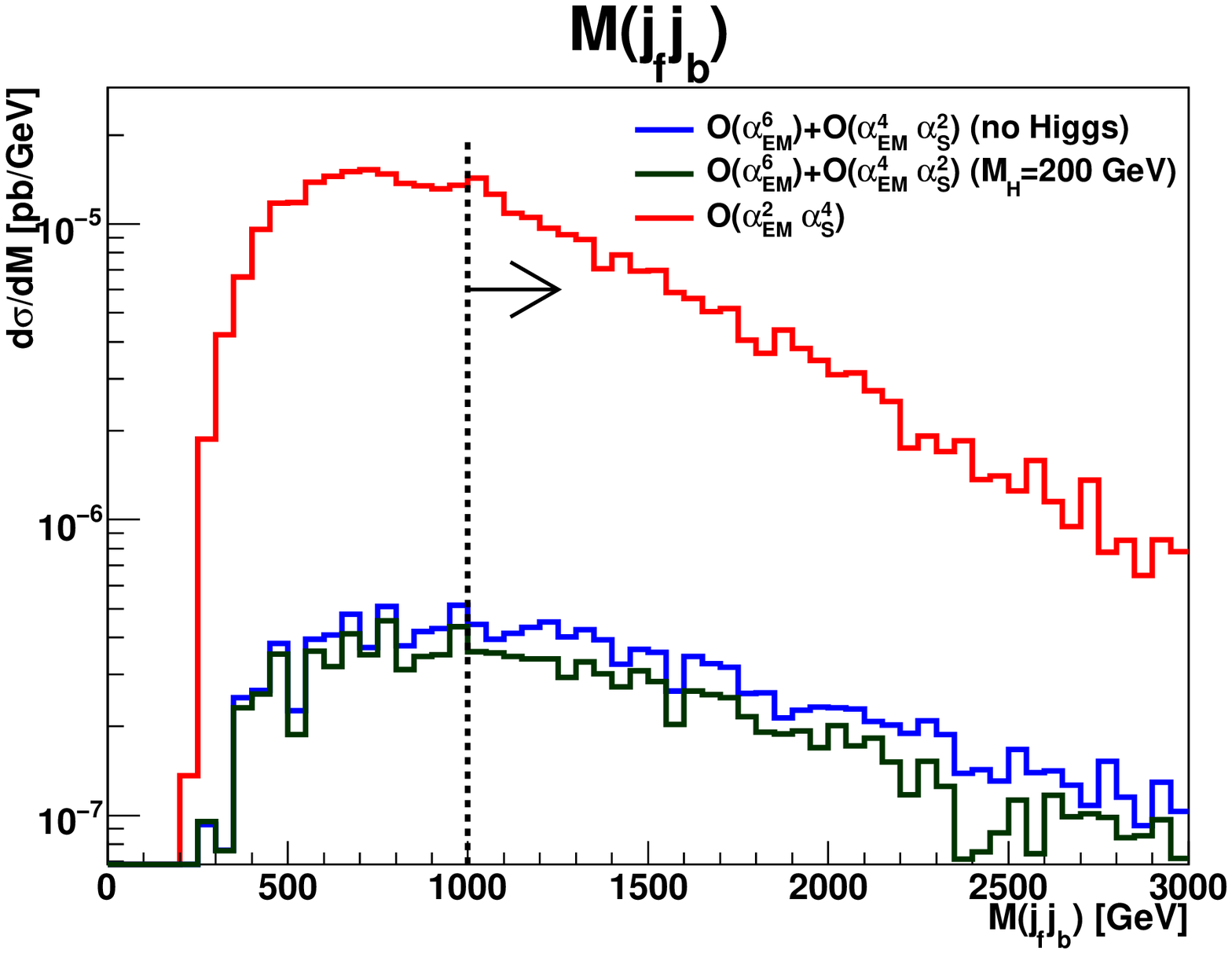}
\hspace*{-0.7cm}
\includegraphics*[width=8.3cm,height=6.2cm]{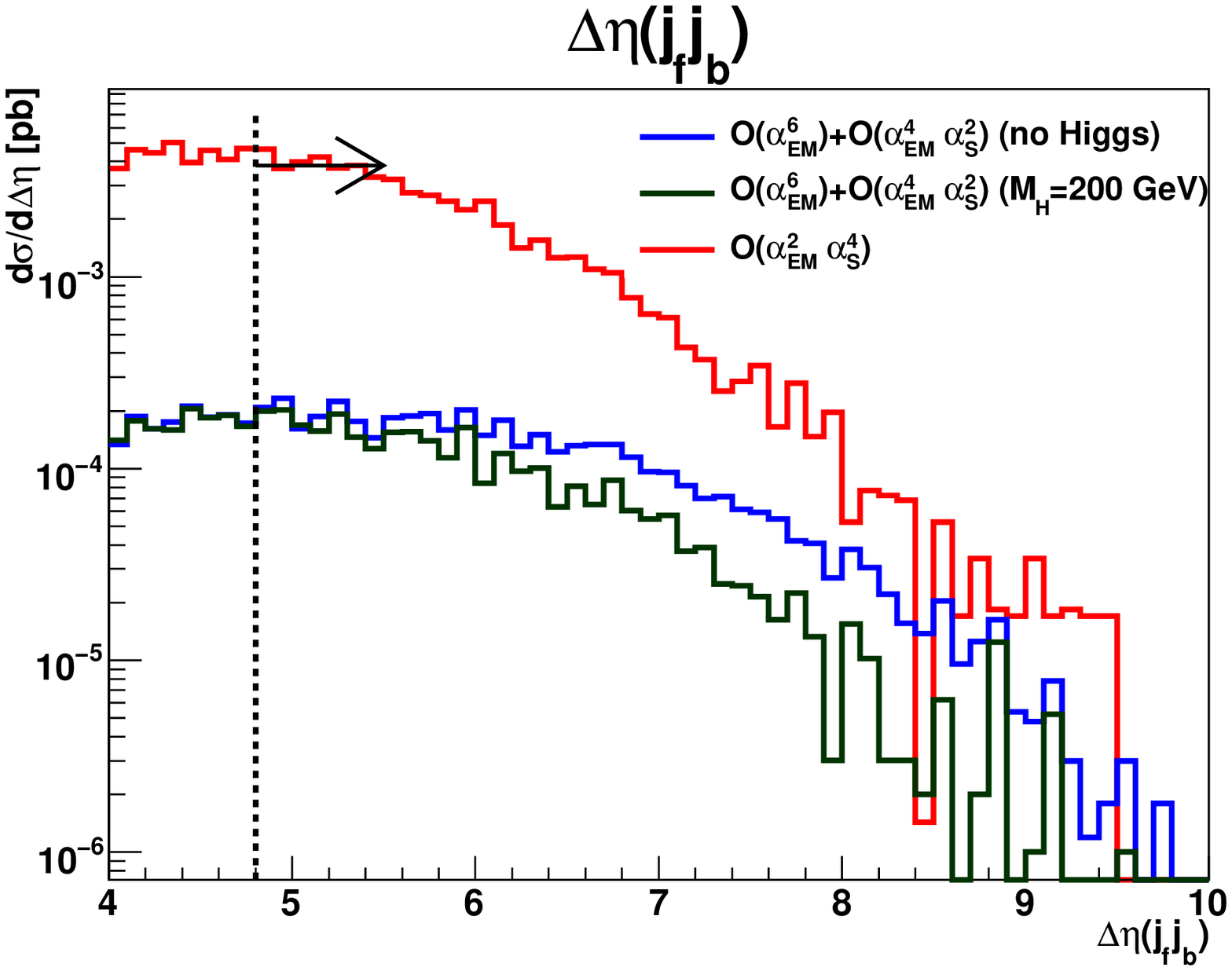}
\hspace*{-3cm}
}
\caption{\textit{Top row:}
distribution of the largest absolute pseudorapidity of the charged leptons
and the corresponding distribution normalized to unit area.
\textit{Bottom row:}distributions of the mass
and separation in pseudorapidity of the two tag jets.
The numbers refer to the
$\mu\mu + 4j$ channel only. The set of cuts in \tbn{tab:cuts_0} are always imposed. The
cuts in \tbn{tab:cuts_1} are applied incrementally. For instance the distribution of
$|\Delta \eta (j_fj_b)|$ at the bottom of the figure includes the additional
cuts $|\eta(\ell^\pm)| < 2.0$ and $M(j_fj_b)>1000 \mbox{ GeV}$. Interferences
between the different perturbative orders are neglected.
}
\label{fig:plots_step2_2}
\end{figure}

\begin{figure}[htb]
\centering
\subfigure{
\hspace*{-2.1cm}
\includegraphics*[width=8.3cm,height=6.2cm]{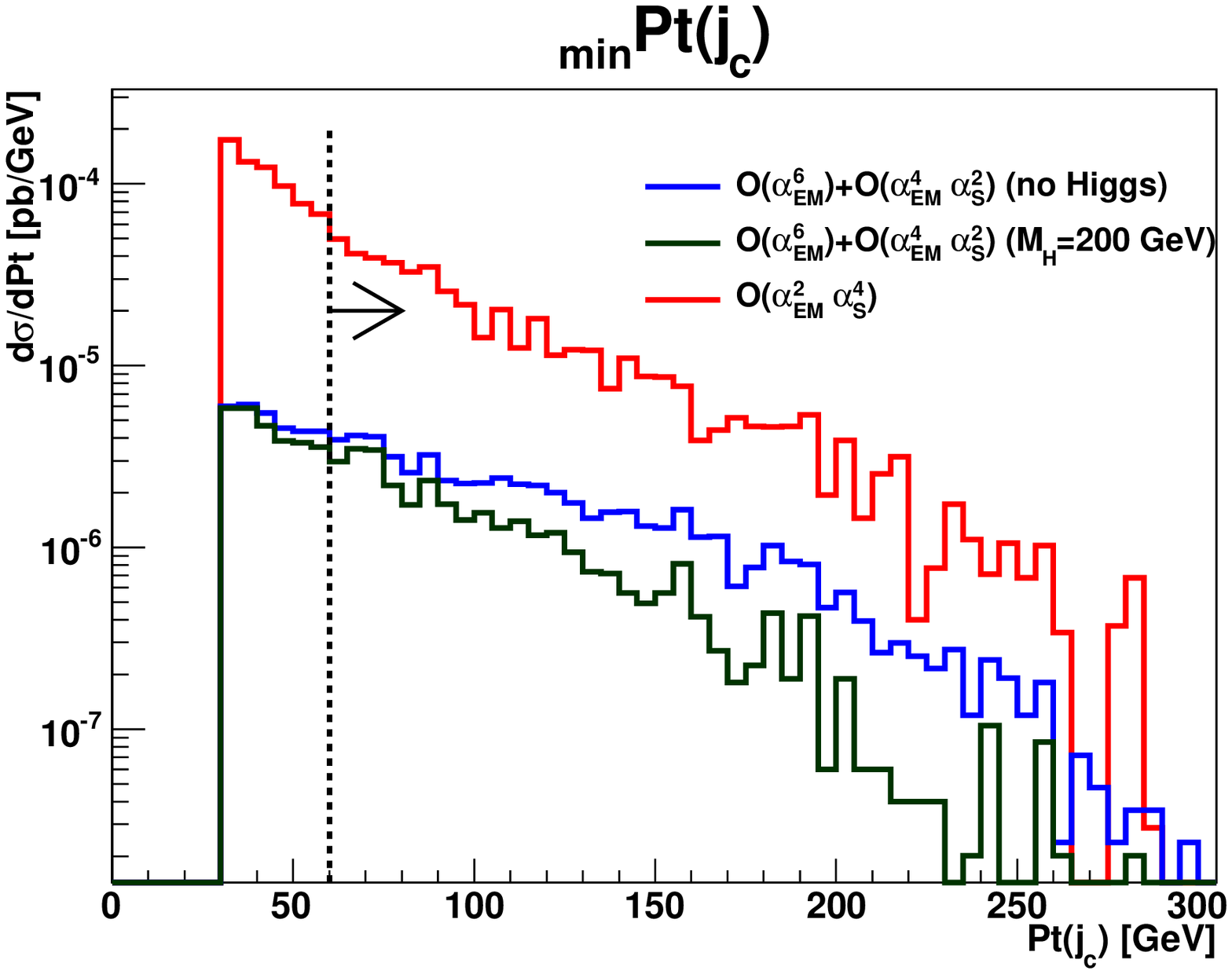}
\hspace*{-0.7cm}
\includegraphics*[width=8.3cm,height=6.2cm]{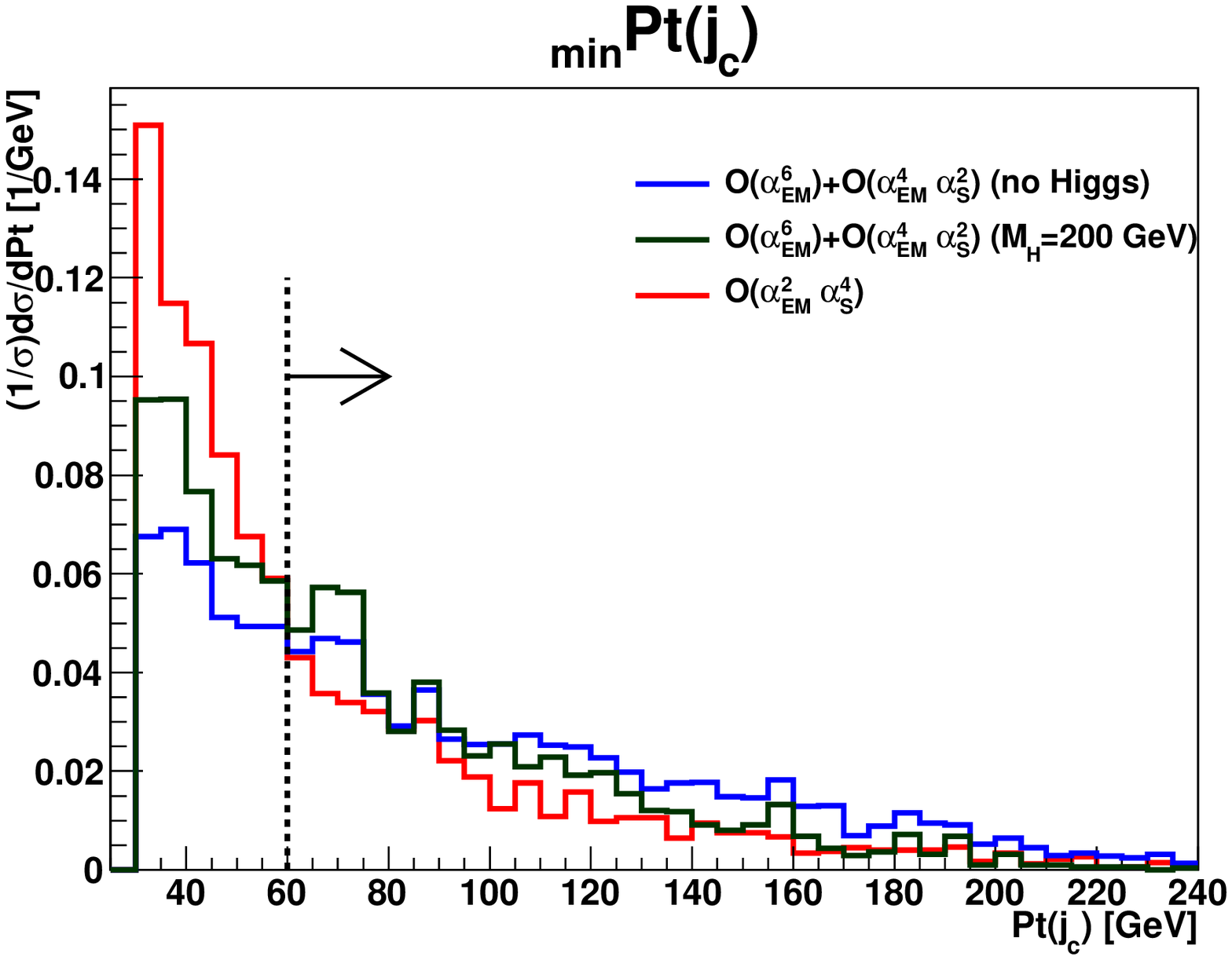}
\hspace*{-3cm}
}
\vspace{-0.4cm}
\subfigure{
\hspace*{-2.1cm}
\includegraphics*[width=8.3cm,height=6.2cm]{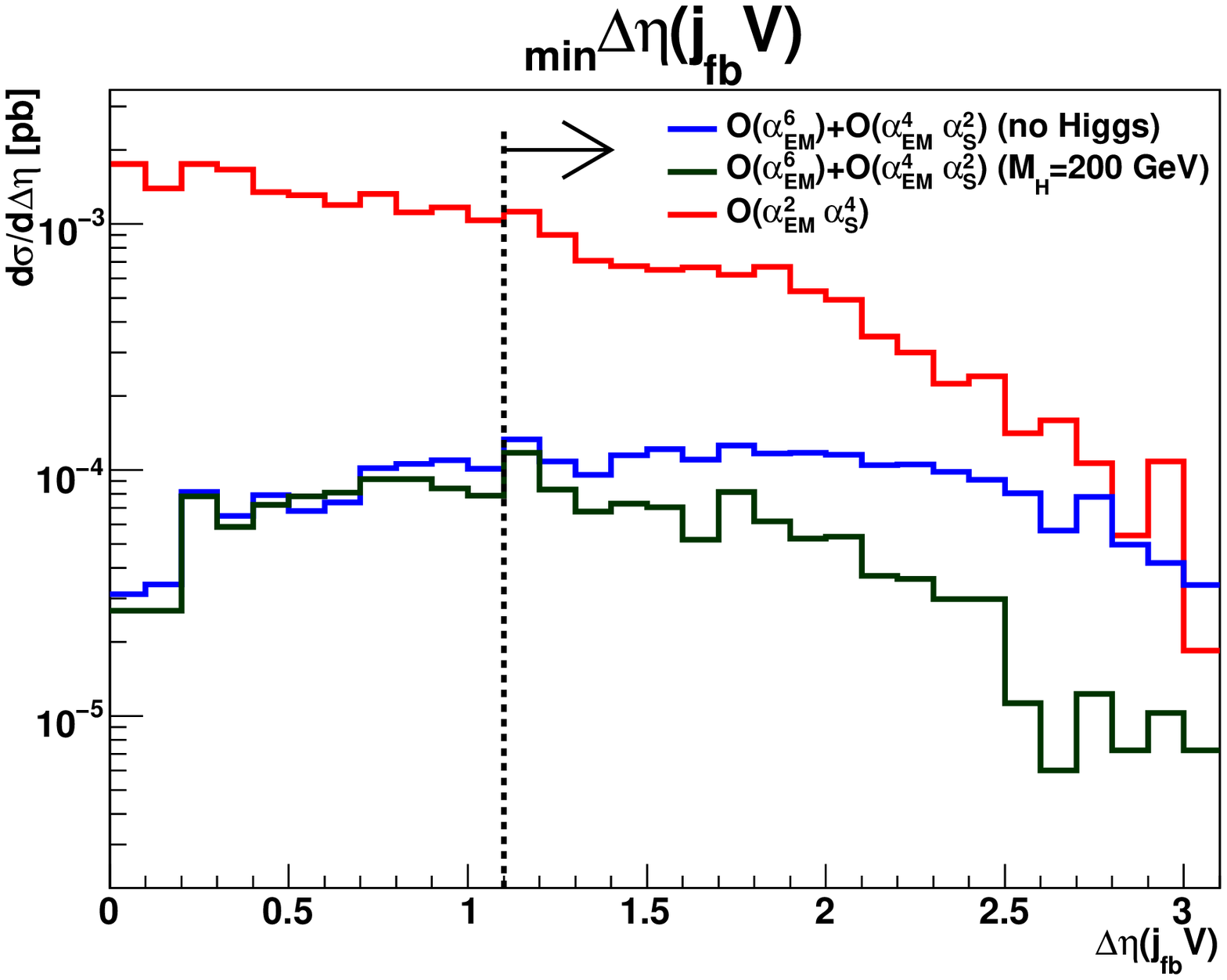}
\hspace*{-0.7cm}
\includegraphics*[width=8.3cm,height=6.2cm]{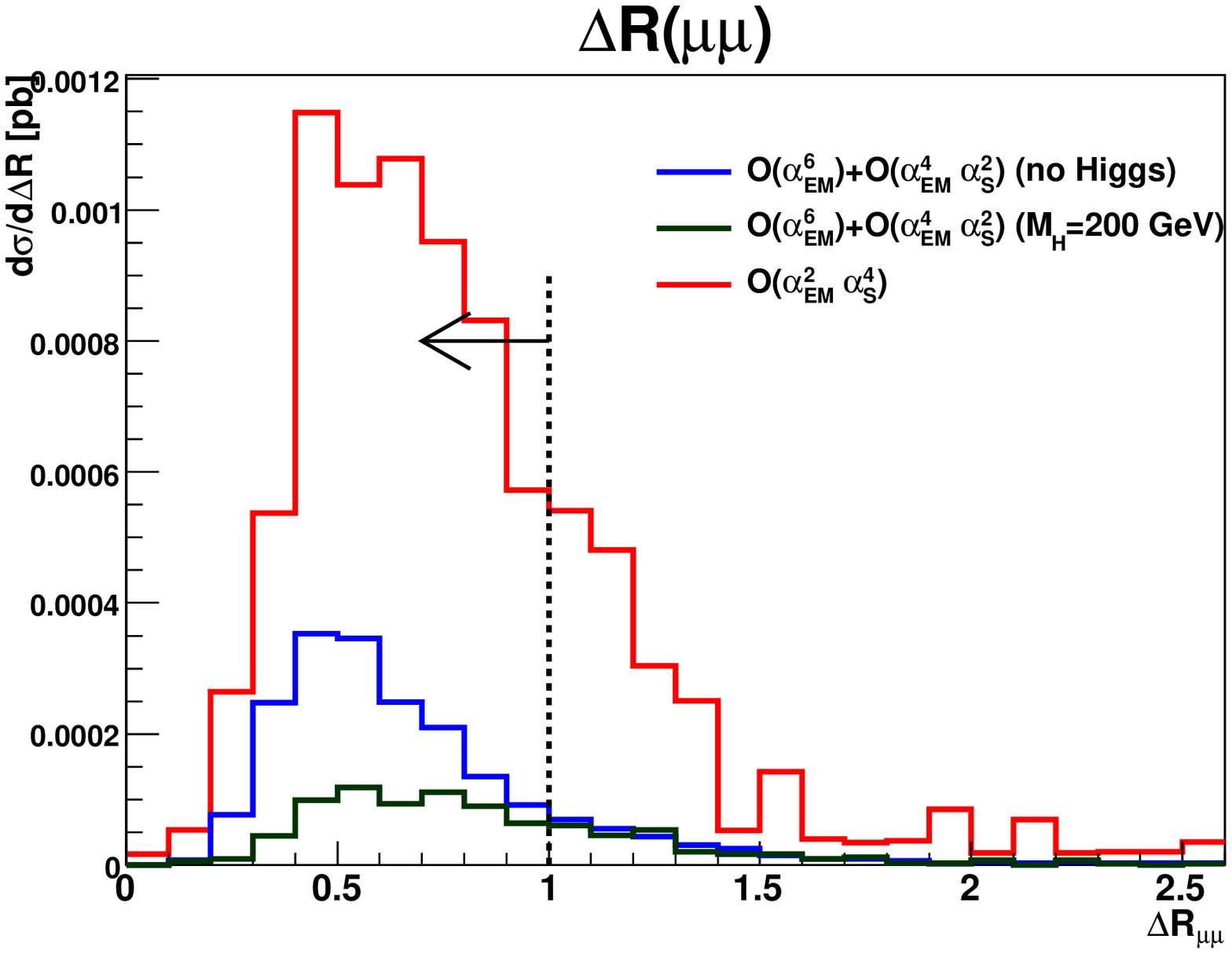}
\hspace*{-3cm}
}
\vspace{0.2cm}
\caption{\textit{Top row:} 
distribution of the smallest transverse momentum of the two central jets 
and the corresponding distribution normalized to unit area.
\textit{Bottom row:} distributions of the smallest separation in pseudorapidity
between the vector bosons and any tag jet
and $|\Delta R|$ separation between charged leptons.
The numbers refer to the $\mu\mu + 4j$ channel only.
The set of cuts in \tbn{tab:cuts_0} are always imposed. The
cuts in \tbn{tab:cuts_1} are applied incrementally. For instance the distribution of
$|\Delta \eta (Vj)|$ includes 
all the additional cuts presented in \fig{fig:plots_step2_2} and in the top row of the
present figure.
Interferences between the two perturbative orders are neglected.
}
\label{fig:plots_step2_3}
\vspace{-0.4cm}
\end{figure}

Therefore, on the generated samples we have applied some additional selection
cuts. They are shown in \tbn{tab:cuts_1} for the $\ell^+ \ell^- + 4j$ channel in the order
in which they have been implemented. The corresponding distributions are
presented in \figs{fig:plots_step2_2}{fig:plots_step2_4} for the SM and Higgsless cases.
In each figure all
previous cuts are applied. The vertical dotted line indicates the value of the
cut and the arrow indicates which part of the events is kept. For the 
distribution of the largest absolute pseudorapidity of the charged leptons
in \fig{fig:plots_step2_2} and the 
distribution of the smallest transverse momentum of the two central jets
in \fig{fig:plots_step2_3} we also present the plots normalized to unit area in
order to show more clearly the differences in shape between the different cases.

The essence of the set of cuts in \tbn{tab:cuts_1} can be easily understood: they amount
to requiring that the two tag jets are highly energetic and well separated both
from each other and from the candidate vector bosons; moreover the two
reconstructed vector bosons are required to be central and have large transverse
momentum.

In \tbn{tab:res_Z4j_1} we present the total cross section with the full set of cuts
in \tbn{tab:cuts_0} and \tbn{tab:cuts_1} as the minimum invariant mass for the $j_c j_c
\ell^+\ell^-$ system is increased between 600 and 1200 GeV.
These results refer to the mass 
window $70 \mbox{ GeV} < M(j_cj_c) < 100 \mbox{ GeV}$. In parentheses the results for the 
$\ordEW + \ordQCD$ 
samples. The last column gives the cross sections for the $\ordQCD$ processes alone;
the reported values are computed with the Higgs mass taken to infinity, they agree
within statistical errors with those obtained for $M_H=200$ GeV.

\begin{table}[htb]
\begin{center}
\begin{tabular}{|c|}
\hline
\textbf{Selection cuts} \\
\hline
$|\eta(\ell^\pm)| < 2.0$ \\
\hline
$M(j_fj_b)>1000 \mbox{ GeV}$ \\
\hline
$|\Delta \eta (j_fj_b)| > 4.8$ \\
\hline
$p_T(j_c) > 60 \mbox{ GeV}$ \\
\hline
$|\Delta \eta (Vj)| > 1.1$ \\
\hline
$\Delta R(\ell^+\ell^-) < 1.0$ \\
\hline
$p_T(\ell^+\ell^-) > 200 \mbox{ GeV}$ \\
\hline
$p_T(j_c j_c) > 200 \mbox{ GeV}$ \\
\hline
$\mbox{Max}|\eta(j)| > 2.8$ \\
\hline
\end{tabular}
\caption{Additional cuts for the $\ell^+ \ell^- + 4j$ channel. 
} 
\label{tab:cuts_1}
\end{center}
\end{table}

\begin{figure}[htb]
\centering
\vspace*{-1.5cm}
\subfigure{
\hspace*{-2.1cm}
\includegraphics*[width=8.3cm,height=6.2cm]{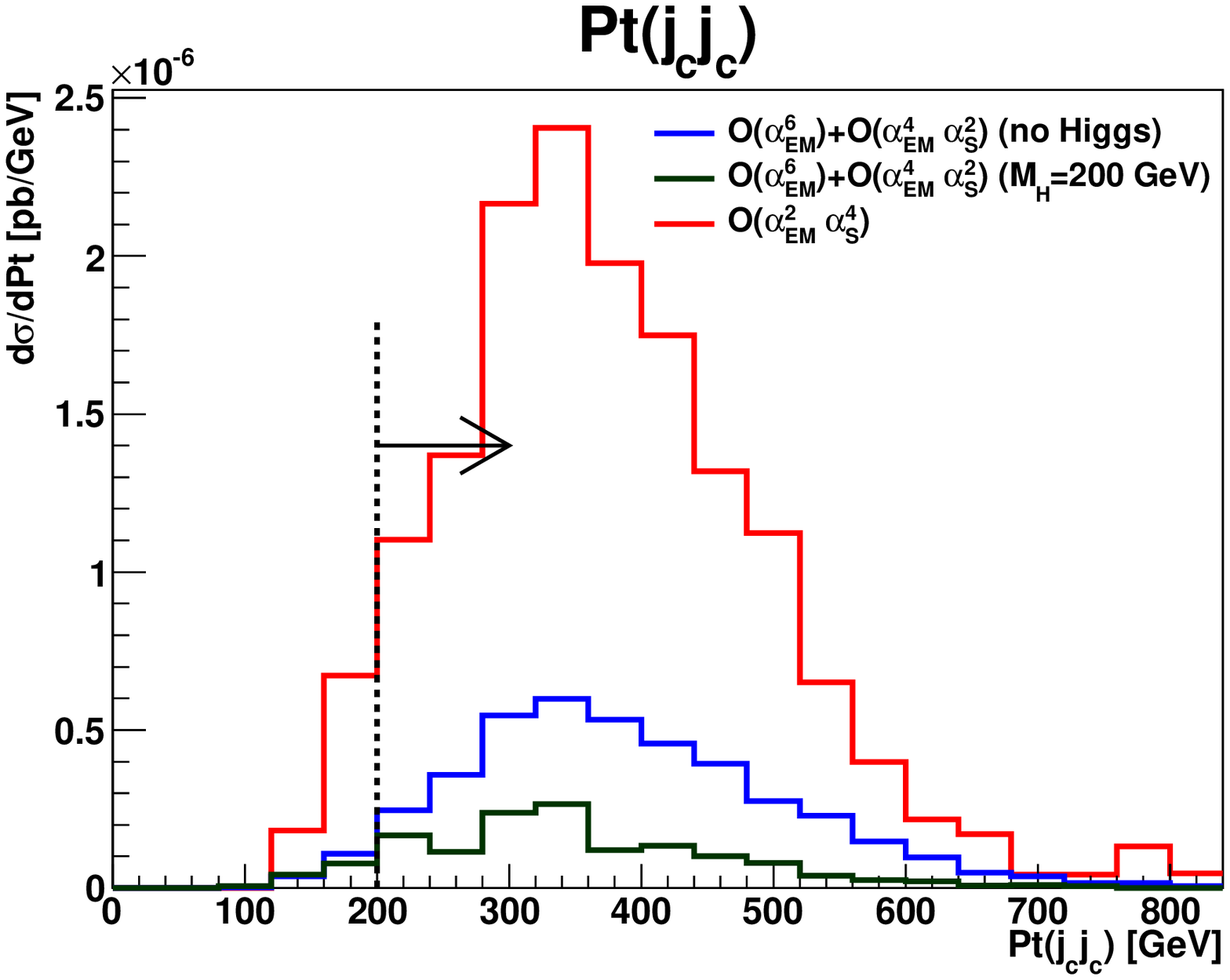}
\hspace*{-0.7cm}
\includegraphics*[width=8.3cm,height=6.2cm]{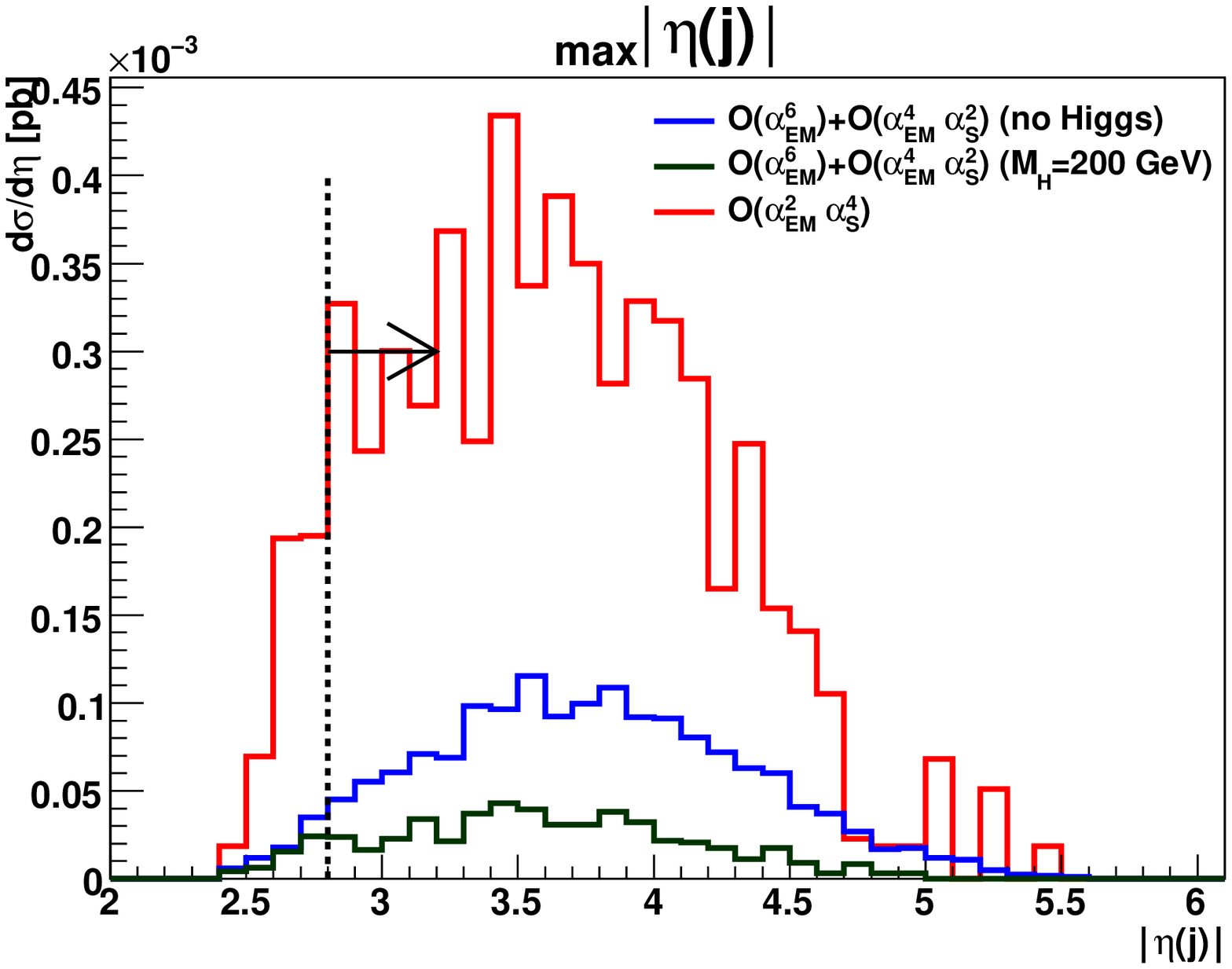}
\hspace*{-3cm}
}
\vspace{0.2cm}
\caption{ 
Distributions of the combined transverse momentum
of the two central jets and the largest absolute pseudorapidity of all jets. The
numbers refer to the $\mu\mu + 4j$ channel only. The set of cuts in \tbn{tab:cuts_0} are
always imposed. The cuts in \tbn{tab:cuts_1} are applied incrementally.
Interferences between the two perturbative orders are
neglected.}
\label{fig:plots_step2_4}
\vspace{-0.4cm}
\end{figure}

For both the Higgsless and SILH cases we have computed the probability of what
we call, for brevity and with a slight abuse of language, the BSM hypothesis at
95\% SM exclusion limit, namely the probability that, assuming the absence of a
Higgs boson or that the SILH model correctly describes nature, the result of an
experimental outcome, with a given luminosity, has a chance of less than 5\% in
the Standard Model (PBSM@95\%CL). For this we proceed as described in
Ref.~\cite{Ballestrero:2008gf},
which we briefly summarize for convenience. We define the background $B$ as the
expected yield of the $\ordQCDsq$ $Z + 4j$ processes and the signal $S$ as the
expected number of events from all $\ordEW$ and $\ordQCD$ processes,
both restricted to the mass 
interval $70 \mbox{ GeV} < M(j_cj_c) < 100 \mbox{ GeV}$ and satisfying all cuts
in \tbn{tab:cuts_0} and \tbn{tab:cuts_1}.
These contributions correspond respectively to the representative diagrams of
\fig{fig:diag_as4} ($B$) and \figs{VV-diag}{fig:VBSbckgr_QCD} ($S$).
In other words $S$ corresponds to the events which produce a resonant boson peak
above the flat background due to $V + 4j$ in the mass distribution of the two
central jets.
$B$ and $S$ are considered as random variables
representing the number of background and signal events for
a possible experimental outcome. $\overline{B}$ and $\overline{S}$ are the
corresponding average values which will be taken equal to the predictions of our
simulation for a luminosity $L = 200 \mbox{ fb}^{-1}$ and summing over the
electron and muon channels. We take into account the statistical uncertainty of
both $B$ and $S$ assuming a standard Poisson distribution with average
$\overline{B}$ and $\overline{S}$ respectively. The predicted signal cross
section is also affected by theoretical uncertainties, so the parameter
$\overline{S}$ is itself subject to fluctuations. For $S$ we assume, in addition
to the statistical fluctuations, a theoretical error defined as a flat
distribution in the window $\overline{S} \pm 30\%$ which, in our opinion, is a
reasonable choice to account for both pdf's and scale uncertainties for the
signal. The processes we are interested in require center of mass energies of
the order of the TeV and therefore involve rather large--$x$ quarks, $x \approx
10^{-1} \div 10^{-2}$ at a typical scale $Q$ of about 100 GeV. In this region
the uncertainty due to the parton distribution functions is of the order of
5\% \cite{Martin:2002aw,Martin:2003sk}. As already stated, QCD corrections are in the range of 10\% and, as
a consequence theoretical uncertainties are expected to be well within this
order of magnitude. Only statistical fluctuations have been taken into
consideration in the case of $B$. This is motivated by the fact that the
background is likely to be well measured experimentally from the region outside
the signal peak, so that the theoretical error on $Z + 4j$ is not expected to be
an issue at the time when real data analysis will be performed. We define the
test statistics \cite{Cowan} $D$ using the following prescription,

\be
\label{ed:discriminant}
D = B + S -\overline{B}
\ee

\begin{figure}[htb]
\centering
\includegraphics*[width=12.3cm]{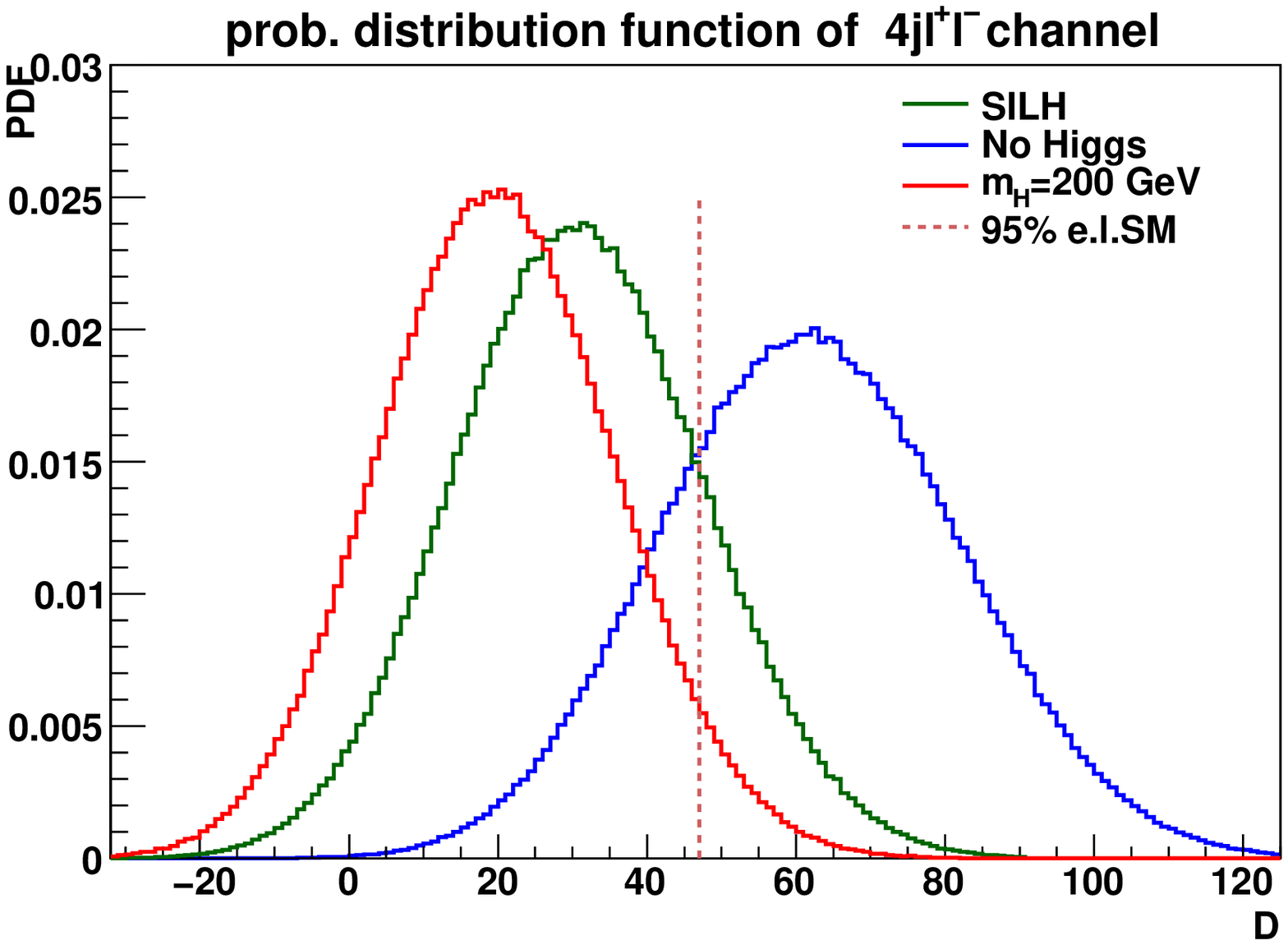}
\caption{
Distributions of the discriminant \eqn{ed:discriminant}
for the $\ell^+ \ell^- + 4j$($\ell = \mu ,\, e$)
channel for $L = 200 \mbox{ fb}^{-1}$ and $M_{cut} = 600 \mbox{ GeV}$ 
with the full set of cuts in \tbn{tab:cuts_0} and \tbn{tab:cuts_1}.
The red curve refers to a Higgs of 200 GeV
while the green one refers to the SILH model and the blue one to the no--Higgs
case. The dotted vertical line in the plot marks the 95\% exclusion limit for
the SM predictions.}
\label{fig:discriminant:Z+4j}
\end{figure}

$D$ represents the actual number of events found in the peak
taking into account statistical fluctuations.
The probability distribution of $D$ for the three scenarios is reported in
\fig{fig:discriminant:Z+4j} for $M_{cut} = 600 \mbox{ GeV}$
with the full set of cuts in \tbn{tab:cuts_0} and \tbn{tab:cuts_1} and under the constraint
$70 \mbox{ GeV} < M(j_cj_c) < 100 \mbox{ GeV}$.
The red curve refers to a Higgs of 200 GeV
while the green one refers to the SILH model and the blue one to the no--Higgs
case. The dotted vertical line in the plot marks the 95\% exclusion limit for
the SM predictions.
We assume a luminosity $L
= 200 \mbox{ fb}^{-1}$ and sum over the $\mu\mu$ and $ee$ final states.

\begin{table}[htb]
\begin{tabular}{|p{0.07\textwidth}|p{0.15\textwidth}|p{0.07\textwidth}|p{0.15\textwidth}|p{0.07\textwidth}|p{0.18\textwidth}||p{0.11\textwidth}|}
\hline
$M_{cut}$ 	& \multicolumn{2}{|c|}{no Higgs}& \multicolumn{2}{|c|}{SILH} 	& $M_H=200$ GeV   & $\ordQCD$\\ 
		\cline{2-7}
 (GeV) 		& \centering $\sigma$(fb) & \centering PBSM & \centering $\sigma$(fb) & \centering PBSM	& 
                                                                                 \parbox[t]{0.2\textwidth}{\centering $\sigma$(fb)} 
                                                                                & \parbox[t]{0.1\textwidth}{\centering $\sigma$(fb)}\\
\hline
600		& 0.705(0.154)	& 77.1\% & 0.625(0.076)  & 16.8\% & 0.595(0.0484) & 0.022 \\
800		& 0.415(0.102)	& 75.3\% & 0.358(0.0457) & 16.1\% & 0.337(0.025)  & 0.010 \\
1000		& 0.209(0.06)   & 65.8\% & 0.174(0.0245) & 12.3\% & 0.162(0.0129) & 0.005 \\
1200		& 0.117(0.0328) & 44.7\% & 0.098(0.0141) & 9.91\% & 0.092(0.0077) & 0.003 \\
\hline
\end{tabular}
\caption{
Total cross section for the $\mu^+ \mu^- + 4j$ channel, with the full set of cuts in \tbn{tab:cuts_0} and 
\tbn{tab:cuts_1}, as a function of the minimum invariant mass $M_{cut}$ for the 
$j_c j_c \mu^+ \mu^-$ system, in the mass 
window $70 \mbox{ GeV} < M(j_cj_c) < 100 \mbox{ GeV}$. In parentheses the results for the 
$\ordEW + \ordQCD$ 
samples. The last column gives the cross sections for the $\ordQCD$ processes alone;
the reported values are computed with the Higgs mass taken to infinity, they agree
within statistical errors with those obtained for $M_H=200$ GeV.
The PBSM probabilities refer to a luminosity of $L = 200 \mbox{ fb}^{-1}$
and to the sum of the 
electron and muon channels.
}
\label{tab:res_Z4j_1}
\end{table}

As reported in \tbn{tab:res_Z4j_1},
the probability of an experiment to find a result incompatible with the SM at
95\%CL assuming that the Higgsless model is realized in Nature is of the order
of 77\% for $M_{cut} = 600 \mbox{ GeV}$ and decreases to about 45\% for $M_{cut}
= 1200 \mbox{ GeV}$. At first sight this is surprising since the difference
between the SM and the Higgsless case increases with $M_{cut}$ if only the
$\ordEW + \ordQCD$ processes are considered. In fact if for instance one
considers the $\ordEW + \ordQCD$ ratio of the Higgsless to the SM cross
sections, this can be seen to be larger at higher $M_{cut}$. This behaviour can
however be understood qualitatively in the following way. We can neglect as a
first approximation the effect of theory errors which only affects the $\ordEW +
\ordQCD$ component. Then, the separation, in terms of events for a $L = 200
\mbox{ fb}^{-1}$ luminosity and two lepton channels, between the no--Higgs and
the light Higgs case amounts to about 2.5 times the standard deviation of the
two distributions for $M_{cut} = 600 \mbox{ GeV}$. For $M_{cut} = 1200 \mbox{
GeV}$ however the same separation is about equal to the corresponding standard
deviation and therefore the overlap between the two distributions is larger and
the statistical discriminating power is reduced.
In other terms, the increase of the relative statistical error as the total
number of
events diminishes at larger $M_{cut}$ is not compensated by the increase of the
ratio between the number of genuine scattering events in the Higgsless case
compared to the SM one.

For the SILH model the
PBSM@95\%CL is only about 17\% at most, for $M_{cut} = 600 \mbox{ GeV}$.

For $L
= 200 \mbox{ fb}^{-1}$ and summing over the $\mu\mu$ and $ee$ final states, the
expected total rates are about 280/250/240 events for the Higgsless/SILH/SM case for
$M_{cut} = 600 \mbox{ GeV}$ and 47/40/37 for $M_{cut} = 1200 \mbox{ GeV}$.
The $\ordQCDsq$ background, 
in the mass window $70 \mbox{ GeV} < M(j_cj_c) < 100 \mbox{ GeV}$ we
are focusing on, yields about 220 events for $M_{cut} = 600 \mbox{ GeV}$ and
33 events for $M_{cut} = 1200 \mbox{ GeV}$.


\section{The $3\ell\nu + 2j$ channel}
\label{sec:3l+2j}

The total cross section for the $3\ell\nu + 2j$ channel with the acceptance cuts
in \tbn{tab:cuts_0} is presented in \tbn{tab:res_3l_0} as a function of the minimum invariant mass
$M_{cut}$ for the $3\ell\nu$ system. In parentheses the results for the $\ordEW$
processes. 

\begin{table}[htb]
\begin{tabular}{|p{0.08\textwidth}|p{0.27\textwidth}|p{0.27\textwidth}|p{0.26\textwidth}|}
\hline
$M_{cut}$ 	& \centering no Higgs 	& \centering SILH  & 
					\parbox[t]{0.26\textwidth}{\centering $M_H=200$ GeV}   \\ 
		\cline{2-4}
 (GeV) 		& \centering $\sigma$(fb) & \centering $\sigma$(fb) & \parbox[t]{0.26\textwidth}{\centering $\sigma$(fb)} \\
\hline
600		& 0.499(0.242)	& 0.462(0.213)   & 0.452(0.204)	\\
800		& 0.252(0.134)	& 0.227(0.114)   & 0.219(0.106)	\\
1000		& 0.139(0.078)	& 0.122(0.0635)  & 0.117(0.0585)\\
1200		& 0.0815(0.0476)& 0.071(0.0374)  & 0.0675(0.0339) \\
\hline
\end{tabular}
\caption{Total cross section for the $3\mu\nu + 2j$ and $2\mu e \nu + 2j$
channels after acceptance cuts, \tbn{tab:cuts_0}.
In parentheses the results for the $\ordEW$ 
sample.
}
\label{tab:res_3l_0}
\end{table}

\begin{figure}[htb]
\centering
\vspace*{-1.5cm}
\subfigure{
\hspace*{-2.1cm}
\includegraphics*[width=8.3cm,height=6.2cm]{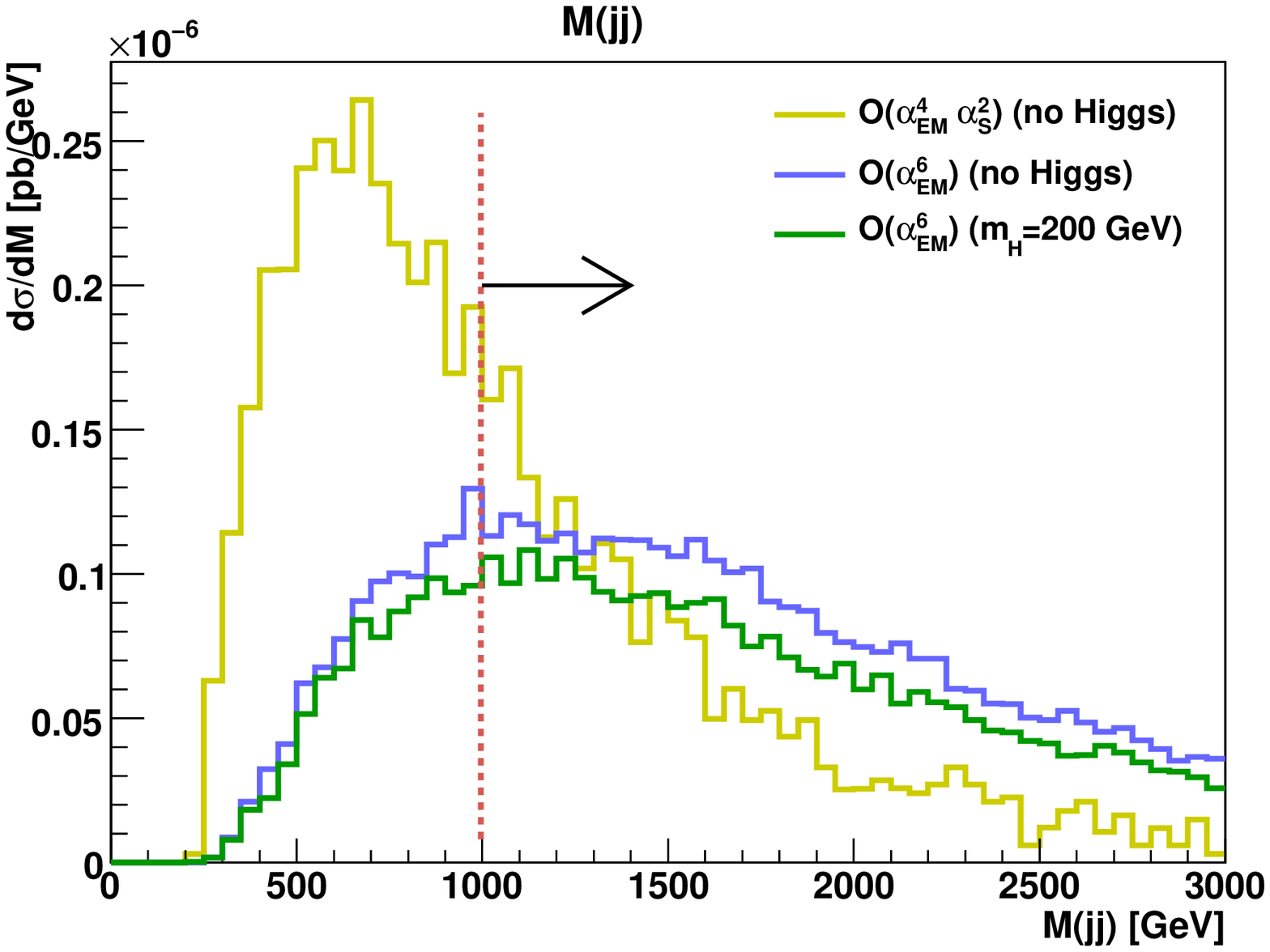}
\hspace*{-0.7cm}
\includegraphics*[width=8.3cm,height=6.2cm]{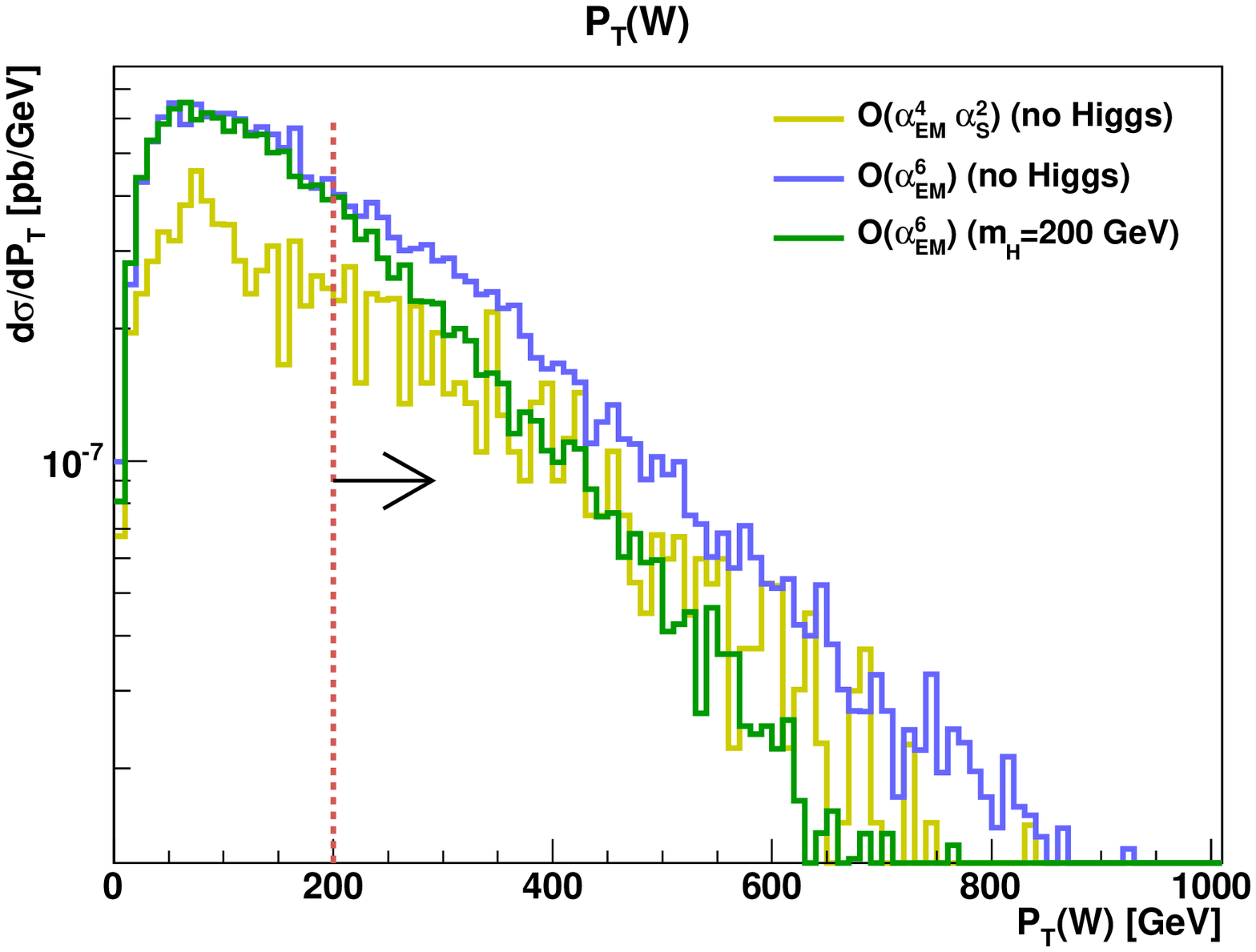}
\hspace*{-3cm}
}
\vspace{-0.4cm}
\subfigure{
\hspace*{-2.1cm}
\includegraphics*[width=8.3cm,height=6.2cm]{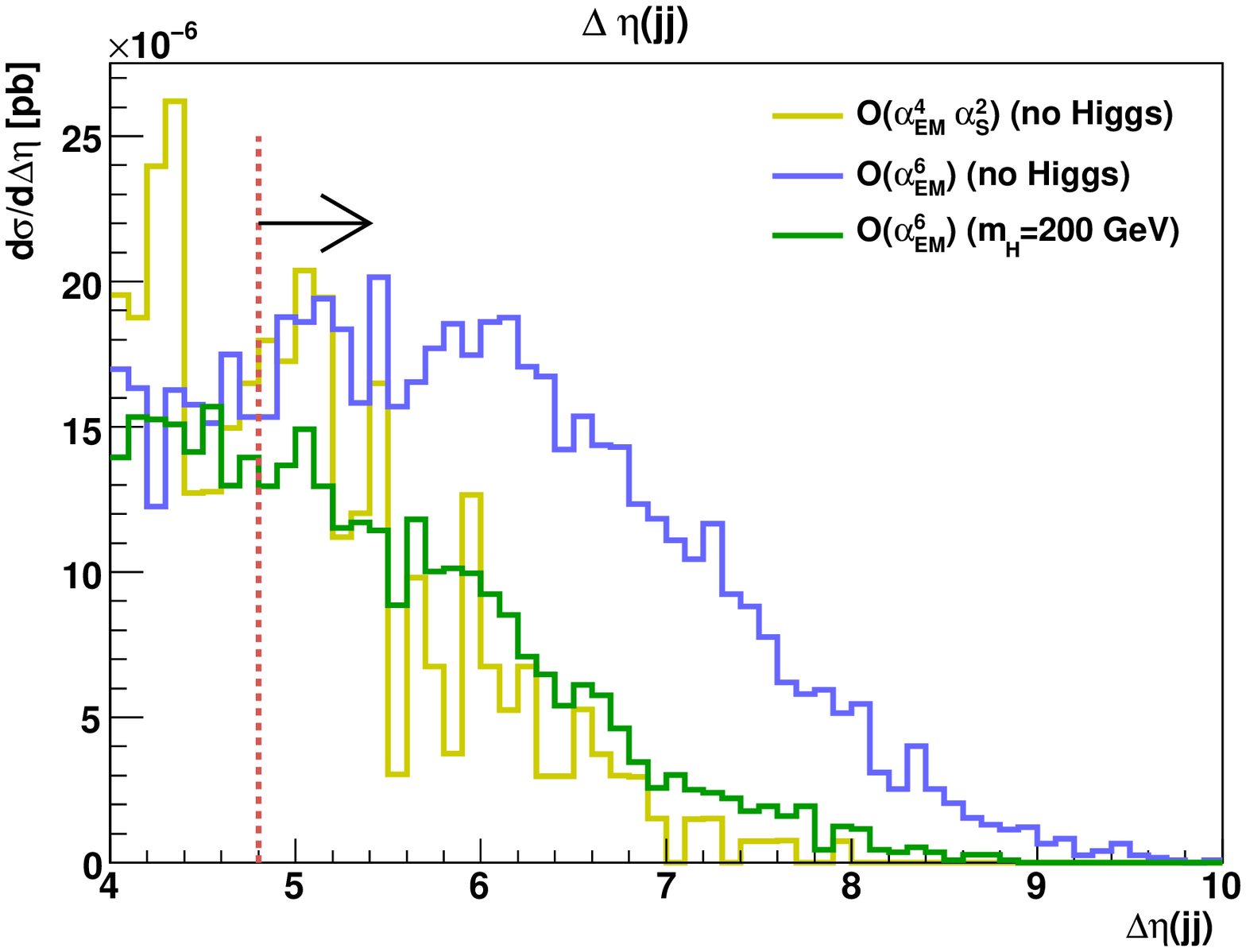}
\hspace*{-0.7cm}
\includegraphics*[width=8.3cm,height=6.2cm]{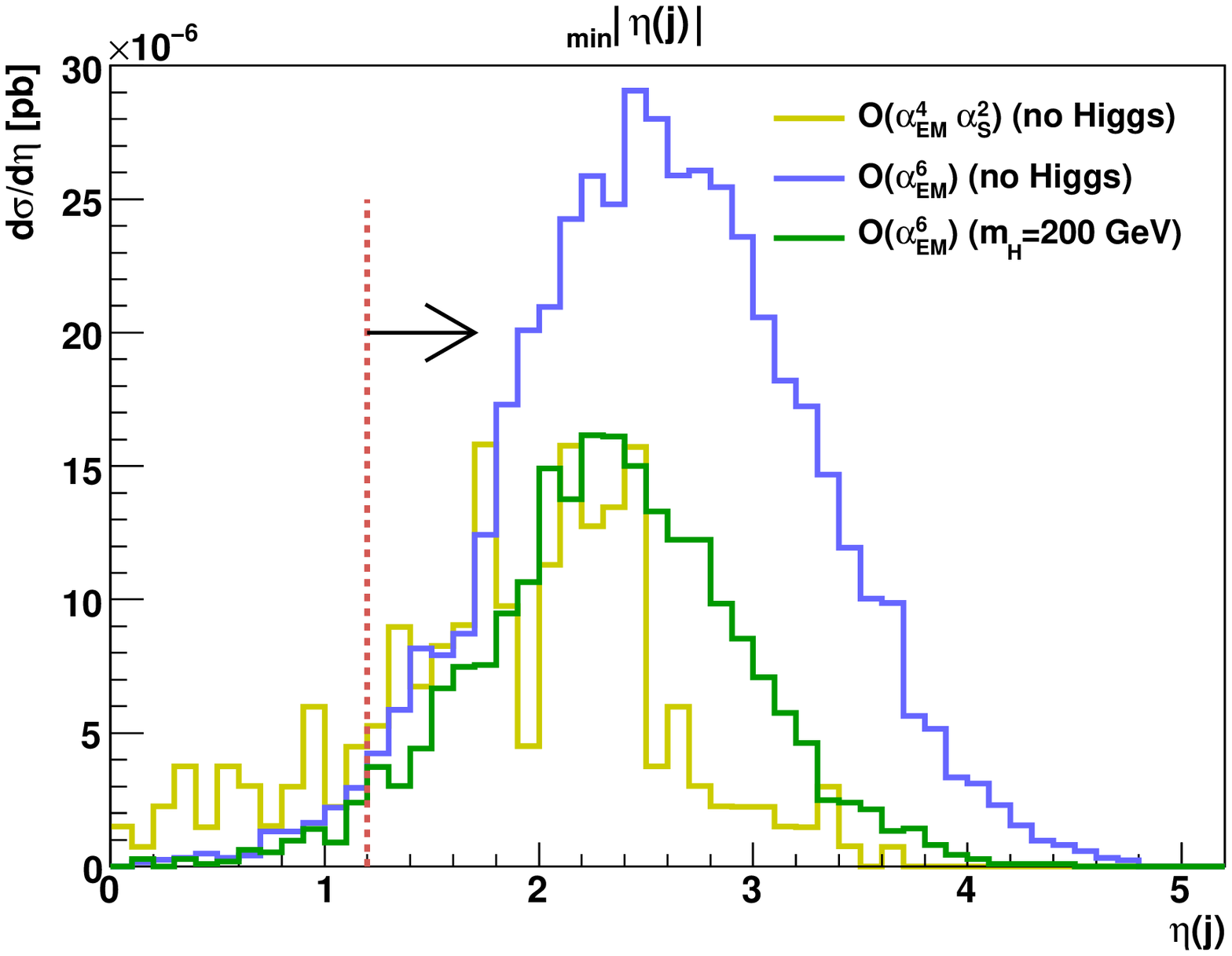}
\hspace*{-3cm}
}
\vspace{0.2cm}
\caption{\textit{Top row:}
distributions of the mass of the two tag jets, the transverse momentum of the
reconstructed $W$.
\textit{Bottom row:} distributions of the separation in pseudorapidity of the
two tag jets and of the largest absolute rapidity of the tag jets.
The numbers refer to the
$\mu\mu\mu + 2j$ and $\mu\mu e + 2j$ channels only.
The set of cuts in \tbn{tab:cuts_0} are
always imposed. The cuts in \tbn{tab:cuts_2} are applied incrementally. For instance the
$|\Delta \eta (j_f j_b)|$
at the bottom of the figure includes the additional cuts
$M(j_fj_b)>1000 \mbox{ GeV}$ and $p_T(\ell^+\ell^-) > 200 \mbox{ GeV}$,
$p_T(\ell\nu) > 200 \mbox{ GeV}$.
Interferences between the two perturbative orders are neglected.}
\label{fig:plots_3l_1}
\vspace{0.4cm}
\end{figure}

\tbn{tab:res_3l_0} shows that the cross section for the $\ordQCD$ processes
is about equal to the cross section for the $\ordEW$
contribution. In the following we will consider the full sample as our signal since 
essentially all events contain a $Z$ and a $W$ boson.
It is possible to improve the discriminating power of the analysis increasing the
fraction of $\ordEW$ events in the event sample since only those are sensitive to the
mechanism of EWSB.
Therefore, on the generated samples we have applied some additional selection
cuts. They are shown in \tbn{tab:cuts_2} in the order in which they have been implemented.
The corresponding distributions are presented in \figsc{fig:plots_3l_1}{fig:plots_3l_2}.
The vertical dotted line indicates the value of the
cut and the arrow indicates which part of the events is kept. The cuts in \tbn{tab:cuts_2},
as in the $\ell^+ \ell^- + 4j$ case, force the two tag jets to be highly
energetic and well separated and the two (reconstructed) vector bosons to be
central and to have large transverse momentum. 

\begin{table}[htb]
\begin{center}
\begin{tabular}{|c|}
\hline
\textbf{Selection cuts} \\
\hline
$M(j_fj_b)>1000 \mbox{ GeV}$ \\
\hline
$p_T(\ell^+\ell^-) > 200 \mbox{ GeV}$ \\
$p_T(\ell\nu) > 200 \mbox{ GeV}$ \\
\hline
$|\Delta \eta (j_fj_b)| > 4.8$ \\
\hline
$|\eta (j_f,j_b)| > 1.2$ \\
\hline
$|\Delta \eta (Vj)| > 1.5$ \\
\hline
$|\eta(\ell^\pm)| < 2.0$ \\
\hline
$M(Vj)> 300 \mbox{ GeV}$ \\
\hline
\end{tabular}
\caption{Additional cuts for the $3\ell\nu + 2j$ channel. 
} 
\label{tab:cuts_2}
\end{center}
\end{table}

The total cross section in
attobarns for the $3\ell\nu + 2j$ channel, with the full set of cuts in \tbn{tab:cuts_0}
and \tbn{tab:cuts_2}, as a function of the minimum invariant mass $M_{cut}$ is shown in
\tbn{tab:res_3l_1}. In parentheses the results for the $\ordEW$ contribution are reported.
The cross section is dominated by the $\ordEW$ contribution. 
The processes including two QCD vertexes
give only a small contribution which decreases sharply at larger $M_{cut}$.

\begin{figure}[htb]
\centering
\subfigure{
\hspace*{-2.1cm}
\includegraphics*[width=8.3cm,height=6.2cm]{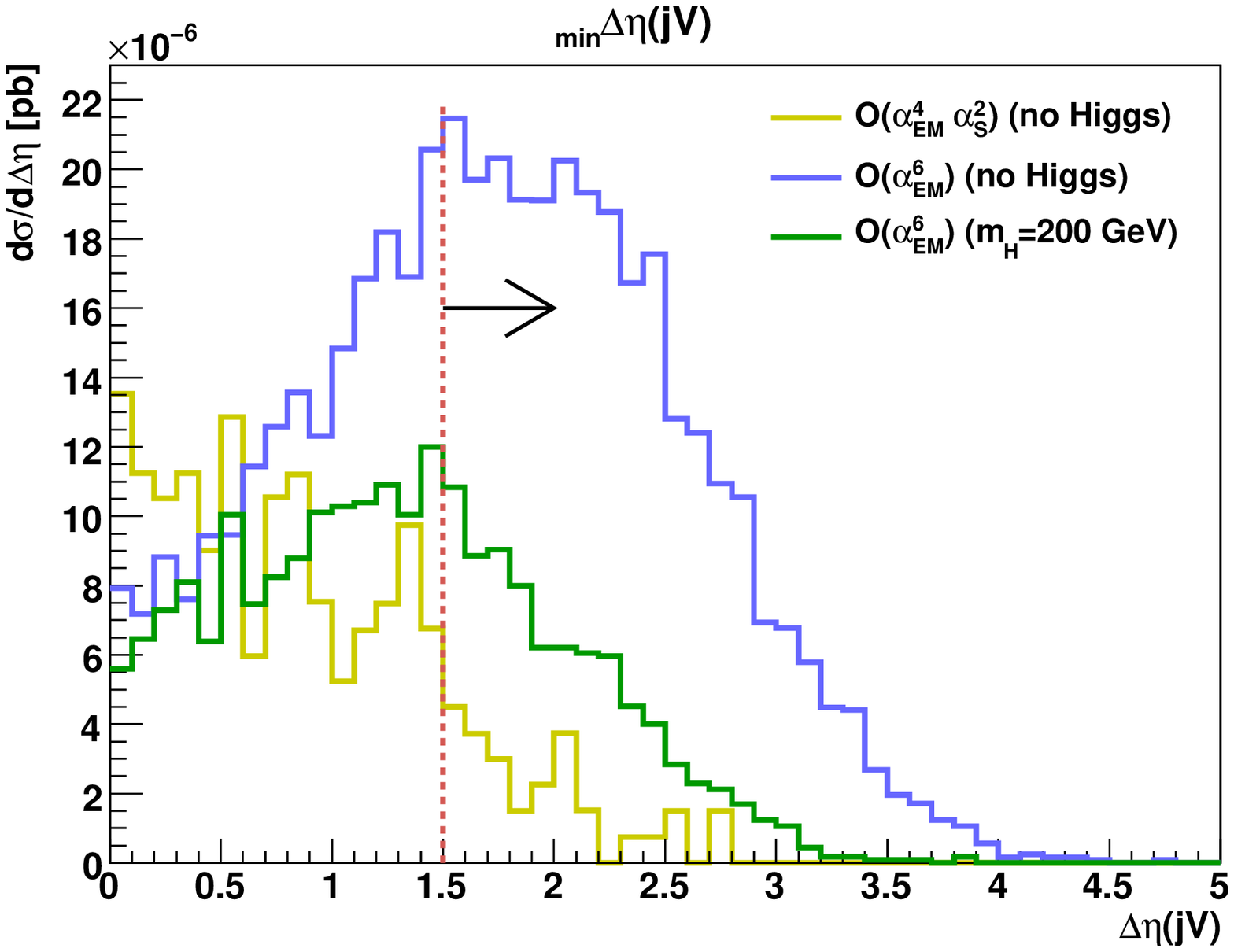}
\hspace*{-0.7cm}
\includegraphics*[width=8.3cm,height=6.2cm]{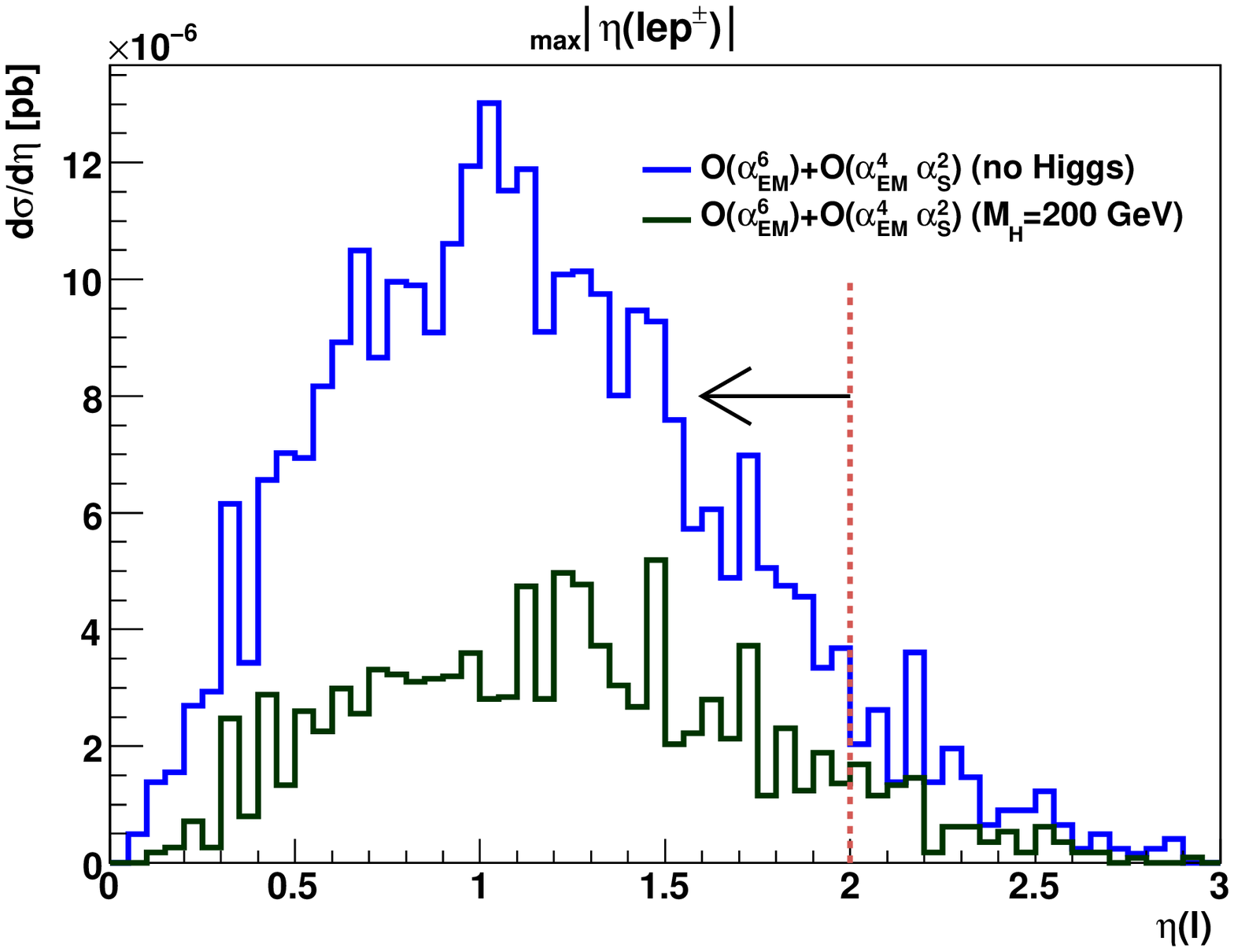}
\hspace*{-3cm}
}
\vspace{-0.4cm}
\subfigure{
\hspace*{-2.1cm}
\includegraphics*[width=8.3cm,height=6.2cm]{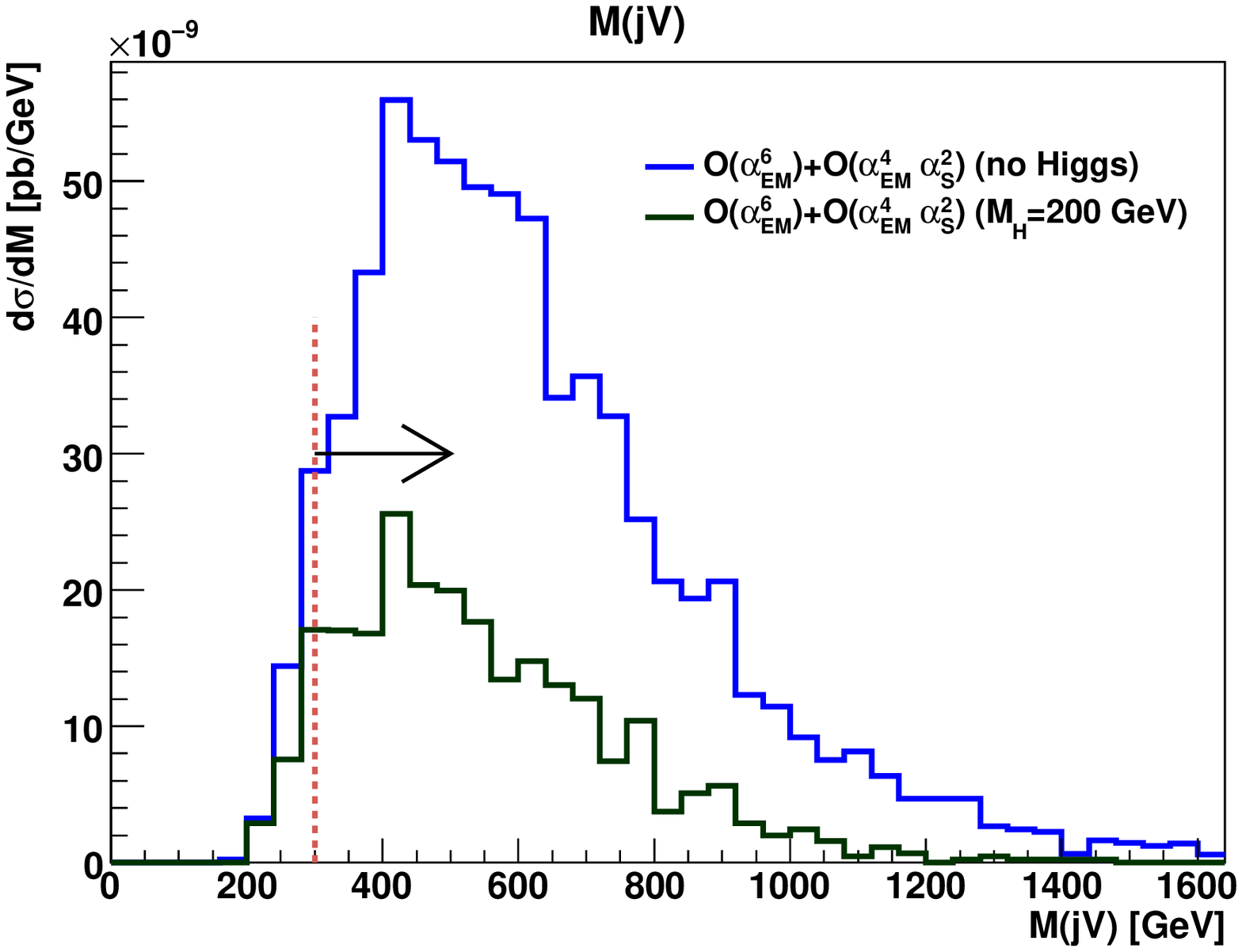}
\hspace*{-3cm}
}
\vspace{0.2cm}
\caption{
Distributions of the smallest separation in
pseudorapidity between the vector bosons and any tag jet, of the smallest absolute
pseudorapidity of any charged lepton and of the smallest invariant mass
of the vector bosons and any tag jet.
The numbers refer to the
$\mu\mu\mu + 2j$ and $\mu\mu e + 2j$ channels only.
The set of cuts in \tbn{tab:cuts_0} are
always imposed. The cuts in \tbn{tab:cuts_2} are applied incrementally.
Interferences between the two perturbative orders are neglected.
}
\label{fig:plots_3l_2}
\end{figure}

\begin{table}[htb]
\begin{tabular}{|p{0.08\textwidth}|p{0.18\textwidth}|p{0.1\textwidth}|p{0.18\textwidth}|p{0.1\textwidth}|p{0.18\textwidth}|}
\hline
$M_{cut}$ 	& \multicolumn{2}{|c|}{no Higgs}& \multicolumn{2}{|c|}{SILH} 	& $M_H=200$ GeV   \\ 
		\cline{2-6}
 (GeV) 		& \centering $\sigma$(ab) & \centering PBSM & \centering $\sigma$(ab) & \centering PBSM	& 
                                                                                 \parbox[t]{0.2\textwidth}{\centering $\sigma$(ab)} \\
\hline
600		& 26.9(24.8) & 80.0\%	& 13.5(11.9) & 19.5\%	& 8.50(6.95) \\
800		& 18.8(17.8) & 84.1\%	& 8.20(7.45) & 24.0\%	& 4.46(3.72) \\
1000		& 12.8(12.6) & 72.9\%	& 5.15(4.79) & 15.9\%	& 2.21(1.83) \\
1200		& 8.65(8.55) & 65.7\%	& 3.08(3.00) & 13.1\%	& 1.19(1.12) \\
\hline
\end{tabular}
\caption{
Total cross section for the $3\mu\nu + 2j$ and $2\mu e \nu + 2j$
channels in attobarns,
with the full set of cuts in \tbn{tab:cuts_0} and 
\tbn{tab:cuts_2}, as a function of the minimum invariant mass $M_{cut}$ for the 
$3\ell\nu$ system. In parentheses the results for the 
$\ordEW$ 
sample. The PBSM probabilities refer to a luminosity of $L = 200 \mbox{ fb}^{-1}$
and to the sum of all electron and muon channels.
}
\label{tab:res_3l_1}
\end{table}

\begin{figure}[htb]
\centering
\includegraphics*[width=12.3cm]{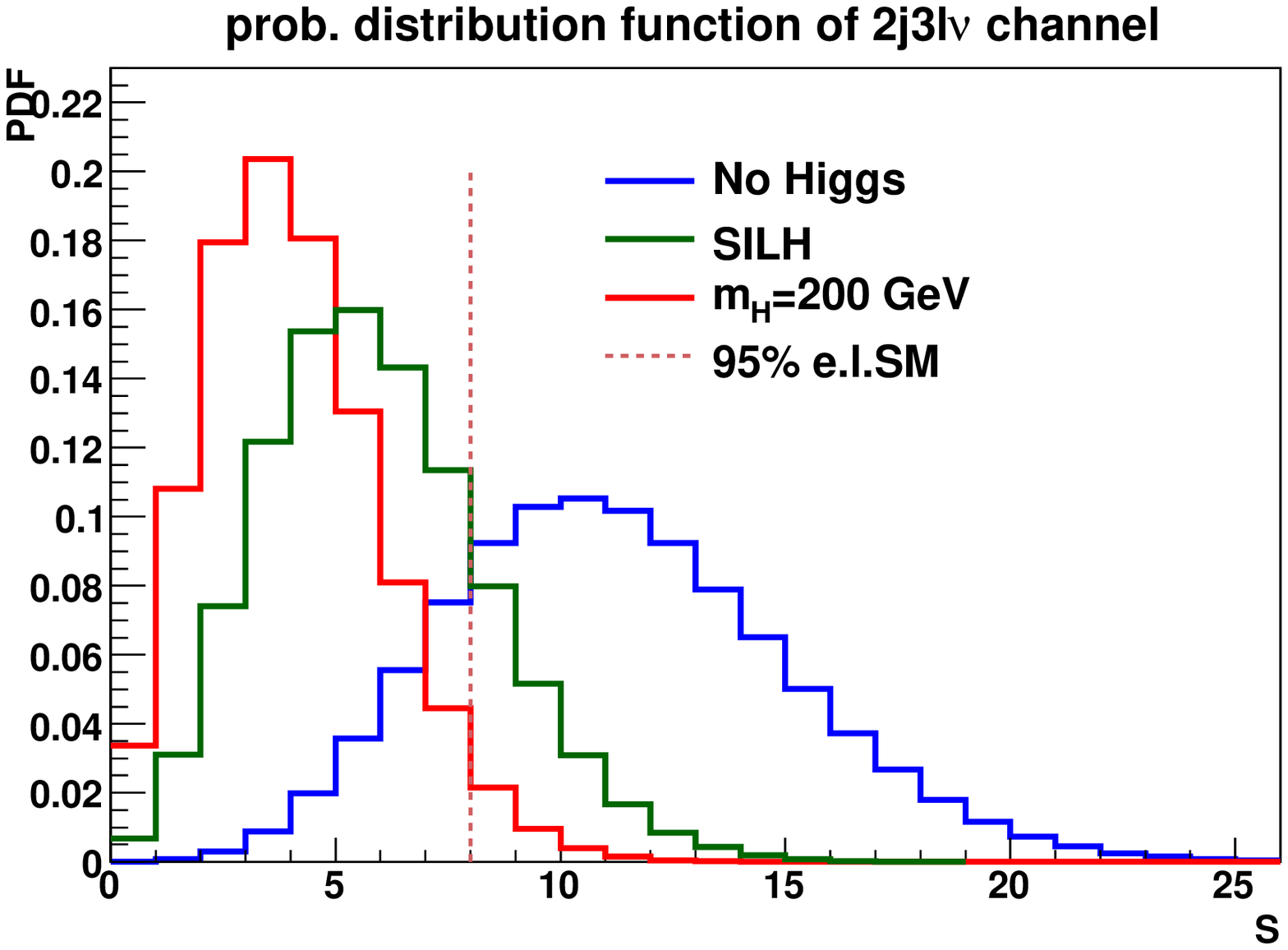}
\caption{
Distributions of the discriminant $D = S$
for the $3\ell\nu + 2j$($\ell = \mu ,\, e$)
channel for $L = 200 \mbox{ fb}^{-1}$ and $M_{cut} = 600 \mbox{ GeV}$. 
The red curve refers to a Higgs of 200 GeV
while the green one refers to the SILH model and the blue one to the no--Higgs
case. The dotted vertical line in the plot marks the 95\% exclusion limit for
the SM predictions.}
\label{fig:discriminant:3l+2j}
\end{figure}

In \tbn{tab:res_3l_1} we also give the PBSM@95\%CL for the two BSM scenarios.
In the present case, since no $\ordQCDsq$ background is present, we use as
discriminant $S$, the sum of the events for all $\ordEW$ and $\ordQCD$ processes
for a luminosity of $L = 200 \mbox{ fb}^{-1}$, summing over the $\mu\mu\mu$,
$\mu\mu e$, $eee$ and $ee\mu$ channels. We take into account the statistical
uncertainty assuming a standard Poisson distribution and we assume a theoretical
error defined as a flat distribution in the window $\overline{S} \pm 30\%$.

The corresponding normalized frequency for the three scenarios is reported in
\fig{fig:discriminant:3l+2j}
for $M_{cut} = 600 \mbox{ GeV}$. The red curve refers to a Higgs of 200 GeV
while the green one refers to the SILH model and the blue one to the no--Higgs
case. The dotted vertical line in the plot marks the 95\% exclusion limit for
the SM predictions.

The probability of an experiment to find a result
incompatible with the SM at 95\%CL assuming that the Higgsless model is realized
in Nature is of the order of 80\% and decreases to about 65\% for $M_{cut} = 1200
\mbox{ GeV}$. Because of the absence of large backgrounds
this channel has a discriminating power which is in fact higher
than the corresponding one for $\ell^+ \ell^- + 4j$ final states.
However the expected rates are quite small. Only about ten events
are expected for all combinations of muons and electrons for a luminosity of $L
= 200 \mbox{ fb}^{-1}$ and $M_{cut} = 600 \mbox{ GeV}$. It might be extremely
difficult to obtain an actual measurement for such a tiny predicted rate once all
the experimental efficiencies are folded in and all sources of isolated leptons
from the underlying event or from jets faking isolated leptons are taken into
account.
Clearly this channel would greatly benefit from a larger luminosity.
The corresponding probabilities for
the SILH model vary between 13\% and 25\%. In the $3\ell\nu + 2j$ channel the
PBSM@95\%CL is somewhat larger at $M_{cut} = 800 \mbox{ GeV}$ than at smaller
value because of the steep decrease of the $\ordQCD$ background at larger
$M_{cut}$.

\eject 

\section{The $\ell\nu + 4j$ channel}
\label{sec:W+4j}

In this section we recall the results presented in Ref.~\cite{Ballestrero:2008gf}
and complete them
with those relative to the SILH model. In addition to the acceptance cuts in
\tbn{tab:cuts_0} we apply the selection cuts in \tbn{tab:cuts_3}.
The corresponding cross sections are shown in \tbn{tab:res_W4j_1}.
These results refer to
the mass window $70 \mbox{ GeV} < M(j_cj_c) < 100 \mbox{ GeV}$ and include
all three perturbative orders. In parentheses
the results for the sum of the $\ordEW$ and $\ordQCD$ processes which we take
as our signal $S$ as for the $\ell^+\ell^- + 4j$ channel.
The last columns gives the cross sections for the $\ordQCD$ processes alone;
the reported values are computed with the Higgs mass taken to infinity, they agree
within statistical errors with those obtained for $M_H=200$ GeV.
The PBSM probabilities are obtained using the procedure
detailed in \sect{sec:Z+4j}.

\begin{table}[htb]
\begin{center}
\begin{tabular}{|c|}
\hline
\textbf{Selection cuts} \\
\hline
$|\eta(\ell^\pm)| < 2.0$ \\
\hline
$M(j_fj_b)>1000 \mbox{ GeV}$ \\
\hline
$|\Delta \eta (j_fj_b)| > 4.8$ \\
\hline
$p_T(j_c) > 70 \mbox{ GeV}$ \\
\hline
$|\Delta \eta (Vj)| > 0.6$ \\
\hline
$p_T(\ell\nu) > 200 \mbox{ GeV}$ \\
\hline
$p_{Tmiss} > 100 \mbox{ GeV}$ \\
\hline
\end{tabular}
\caption{Additional cuts for the $\ell\nu + 4j$ channel. 
} 
\label{tab:cuts_3}
\end{center}
\end{table}

\begin{table}[htb]
\begin{tabular}{|p{0.07\textwidth}|p{0.15\textwidth}|p{0.07\textwidth}|p{0.15\textwidth}|p{0.07\textwidth}|p{0.18\textwidth}||p{0.11\textwidth}|}
\hline
$M_{cut}$ 	& \multicolumn{2}{|c|}{no Higgs}& \multicolumn{2}{|c|}{SILH} 	& $M_H=200$ GeV   & $\ordQCD$\\ 
		\cline{2-7}
 (GeV) 		& \centering $\sigma$(fb) & \centering PBSM & \centering $\sigma$(fb) & \centering PBSM	& 
                                                                                 \parbox[t]{0.2\textwidth}{\centering $\sigma$(fb)} 
                                                                                & \parbox[t]{0.1\textwidth}{\centering $\sigma$(fb)}\\
\hline
600		& 6.07(1.18)	  & 96.5\%	& 5.59(0.704)	   & 35.9\%	& 5.41(0.524) & 0.23 \\
800		& 3.76(0.779)	  & 96.8\%	& 3.40(0.418)	   & 29.2\%	& 3.29(0.309) & 0.13 \\
1000		& 2.26(0.483)	  & 95.4\%	& 2.01(0.227)	   & 19.8\%	& 1.94(0.169) & 0.08 \\
1200		& 1.32(0.263)	  & 83.9\%	& 1.19(0.132)	   & 16.9\%	& 1.15(0.094) & 0.05 \\
\hline
\end{tabular}
\caption{
Total cross section for the $\mu\nu + 4j$ channel, with the full set of cuts in \tbn{tab:cuts_0} and 
\tbn{tab:cuts_3}, as a function of the minimum invariant mass $M_{cut}$ for the 
$j_c j_c \mu\nu$ system, in the mass 
window $70 \mbox{ GeV} < M(j_cj_c) < 100 \mbox{ GeV}$. In parentheses the results for the 
$\ordEW + \ordQCD$ 
samples. The last columns gives the cross sections for the $\ordQCD$ processes alone;
the reported values are computed with the Higgs mass taken to infinity, they agree
within statistical errors with those obtained for $M_H=200$ GeV.
The PBSM probabilities refer to a luminosity of $L = 200 \mbox{ fb}^{-1}$
and to the sum of the 
electron and muon channels.
}
\label{tab:res_W4j_1}
\end{table}

For $M_{cut} = 600 \mbox{ GeV}$ the expected $\ordQCDsq$ background in the mass
window $70 \mbox{ GeV} < M(j_cj_c) < 100 \mbox{ GeV}$ is about 1950 events 
with a luminosity of $L = 200 \mbox{ fb}^{-1}$
and summing over the 
electron and muon channels.
Correspondingly,
209 signal events are expected in the light Higgs SM scenario and 474  in the no
Higgs case. The corresponding prediction in the SILH model is 282 events.

\begin{figure}[htb]
\centering
\includegraphics*[width=12.3cm]{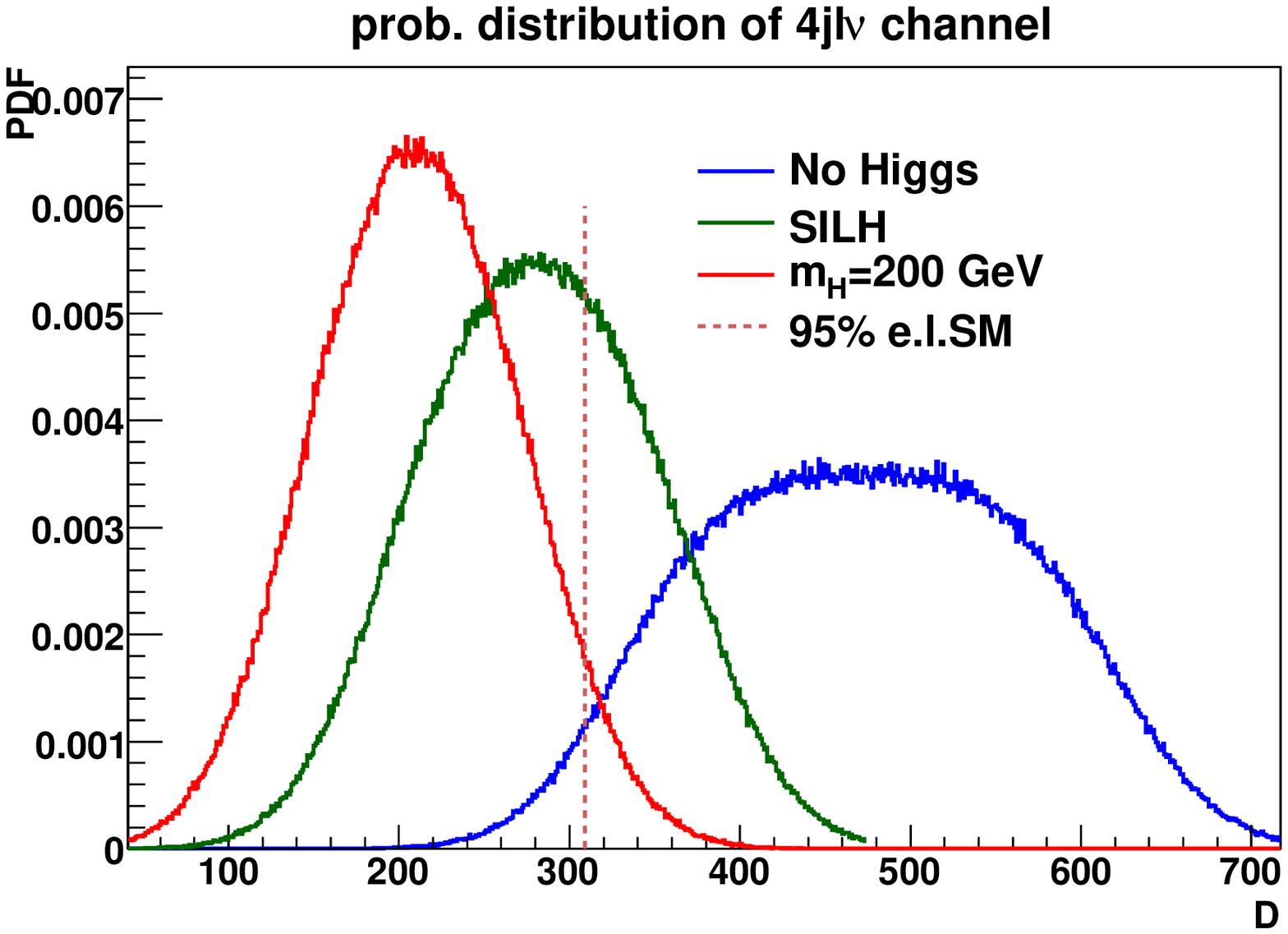}
\caption{
Distributions of the discriminant \eqn{ed:discriminant}
for the $\ell\nu + 4j$($\ell = \mu ,\, e$)
channel for $L = 200 \mbox{ fb}^{-1}$ and $M_{cut} = 600 \mbox{ GeV}$. 
The red curve refers to a Higgs of 200 GeV
while the green one refers to the SILH model and the blue one to the no--Higgs
case. The dotted vertical line in the plot marks the 95\% exclusion limit for
the SM predictions.}
\label{fig:discriminant:W+4j}
\end{figure}

The normalized frequency of the discriminant $D$ for the three scenarios is reported in
\fig{fig:discriminant:W+4j}
for $M_{cut} = 600 \mbox{ GeV}$. The red curve refers to a Higgs of 200 GeV
while the green one refers to the SILH model and the blue one to the no--Higgs
case. The dotted vertical line in the plot marks the 95\% exclusion limit for
the SM predictions.

The $\ell\nu + 4j$ channel, as expected, is the one with the best discriminating
power between the BSM models and the SM. The expected number of events in this
final state is about ten times larger than in the $\ell^+ \ell^- + 4j$ channel.

The probability of an experiment to find a result incompatible with the SM at
95\%CL assuming that the Higgsless model is realized in Nature is of the order
of 96\% for $M_{cut} = 600 \mbox{ GeV}$ and decreases to about
84\% for $M_{cut} = 1200 \mbox{ GeV}$. For the
SILH model the PBSM@95\%CL varies between 35\% and 17\%.


\section{Combining all channels}
\label{sec:combining}

In this section we derive the probability that, assuming that either the
Higgsless scenario or the instance of SILH model we have considered is realized
in Nature, the results of the measurements of the $\ell\nu + 4j$, $3\ell\nu +
2j$ and $\ell^+ \ell^- + 4j$ channels at the LHC yield results which are outside
the 95\% probability region for the SM.
In order to do so, it is convenient to rephrase the
method employed in Ref.~\cite{Ballestrero:2008gf} and in the previous sections
for the case of a single measurement. In the new
language the generalization to a set of several simultaneous measurements will
be obvious. Given two models $A$ and $B$ and the probability distributions of
some physical observable $D$, $P(D|A)$ and $P(D|B)$, the 95\%CL
region for model $A$ can be defined considering the probability ratio

\be
\label{eq:ratio}
R= \frac{P(D|A)}{P(D|A)+P(D|B)} < 1.
\ee

One can consider all the possible $D$ values with
$R>\alpha$ and fix $\alpha$ with the condition that

\be
\label{eq:alphaFix}
\int dD \, P(D|A) \,\theta (R-\alpha) = 95\%
\ee

For distributions like those in \fig{fig:discriminant:Z+4j}, \ref{fig:discriminant:3l+2j}
and \ref{fig:discriminant:W+4j} this procedure would lead for the
SM to the one--sided exclusion regions shown by a vertical dotted line. The
probability for model $B$ to yield a result outside this 95\%CL region for $A$ would
then be

\be
\label{eq:PBSM}
PBSM@95\%CL = \int dD \, P(D|B) \,\theta (\alpha - R) 
\ee

\eqns{eq:ratio}{eq:PBSM} can be easily generalized to a set of observables
$D_1,....,D_N$ introducing the joint probability $P(D_1,....,D_N|A)$ and
transforming the integrals \eqns{eq:alphaFix}{eq:PBSM} to $N$--dimensional ones.
In the present case we have assumed the probabilities to be independent and
defined

\be
\label{eq:N_ProbDef}
P(D_1,....,D_N|A) = \prod_{i=1}^{N} P(D_i|A)
\ee

This method is obviously equivalent to the one, based on the Neyman--Pearson lemma,
of using the likelihood ratio $P(D|A)/P(D|B)$ for discriminating between two different
hypotheses for a multidimensional test statistics \cite{Cowan}.

In \tbn{tab:PBSNM_all} we present the PBSM@95\%CL for the Higgsless and SILH model compared 
to the SM predictions for each channel and for their combination for $M_{cut} = 600 \mbox{ GeV}$.
We assume $L = 200 \mbox{ fb}^{-1}$
and sum over all electron and muon channels.

\begin{table}[htb]
\begin{center}
\begin{tabular}{|c|c|c|}
\hline
channel 	& NOH 	 & SILH \\ 
\hline
$3\ell\nu + 2j$ & 80.0\% & 19.5\% \\
\hline
$\ell^+ \ell^- + 4j$ & 77.1\% & 16.8\% \\
\hline
$\ell\nu + 4j$ & 96.5\%  & 35.9\% \\
\hline
semi-leptonic	& 99.4\% & 41.2\% \\
\hline
all		& 99.9\% & 51.5\% \\
\hline 
\end{tabular}
\caption{PBSM@95\%CL for for the individual channels and for their combination
with the full set of cuts for each channel.
The results labeled semi-leptonic refer to the combination of the $\ell^+ \ell^- + 4j$
and $\ell\nu + 4j$ channels.}
\label{tab:PBSNM_all}
\end{center}
\end{table}

\tbn{tab:PBSNM_all} shows, within the limits of the parton level, lowest-order analysis
presented in this paper, that if no Higgs is present we are essentially certain
that the LHC will obtain combined results
in the $\ell\nu + 4j$, $\ell^+ \ell^- + 4j$ and $3\ell\nu + 2j$ channels which
will lay outside the 95\% CL region of the Standard Model.
If however some version of the SILH models discussed in Ref.~\cite{Giudice:2007fh} is
instead realized the probability of such an outcome drops to about 50\%.

\begin{table}[htb]
\begin{center}
\begin{tabular}{|c|c|c|c|c|c|c|}
\hline
channel 	& \multicolumn{3}{|c|}{NOH} 	 & \multicolumn{3}{|c|}{SILH} \\
\hline
                & $\Delta R = 0.0$ & $\Delta R = 0.3$  & $\Delta R = 0.5$
                & $\Delta R = 0.0$ & $\Delta R = 0.3$  & $\Delta R = 0.5$ \\
\hline
$3\ell\nu + 2j$ &80.0\%  & 80.0\% &80.0\%  &19.5\% & 19.5\% & 19.5\%\\
\hline
$\ell^+ \ell^- + 4j$ & 90.1\% & 77.1\% & 38.5\% & 22.5\%& 16.8\% & 8.5\%\\
\hline
$\ell\nu + 4j$ & 99.9\% & 96.5\%  & 79.9\%& 51.8\% & 35.9\% & 25.1\% \\
\hline
semi-leptonic	& 99.99\% & 99.4\% & 90.1\% & 58.9\% & 41.2\% & 26.8\% \\
\hline
all		& 99.999\% & 99.9\% & 98.7\% & 66.4\% & 51.5\% & 39.8\%\\
\hline 
\end{tabular}
\caption{PBSM@95\%CL for the individual channels and for their combination for
different values of the jet cone separation $\Delta R$. Both the Higgsless case
and the SILH model are considered.}
\label{tab:PBSNM_all_DR}
\end{center}
\end{table}

In \tbn{tab:PBSNM_all_DR} we show how our results depend on the size of the
$\Delta R$ separation cut among jets, comparing our standard choice $\Delta R = 0.3$
with the the requirement of a stronger separation, $\Delta R = 0.5$
and with case in which no cone separation among jet is required, $\Delta R = 0.0$.
In all cases each jet pair is required to have an invariant mass of at least 60 GeV.
As already mentioned in \sect{sec:calculation},
at large transverse momentum, jet pairs with mass comparable to the mass
of electroweak bosons can merge into one single jet when an
angular measure like $\Delta R(jj)$ is adopted for reconstructing jets. Since we
insist on requiring four jets in the final state these events are discarded and
the cross sections are smaller at larger $\Delta R$. However since the $W$'s
transverse momentum distribution is harder in the SILH model and in the Higgsless
case than in the SM the PBSM@95\%CL decreases significantly at larger values of 
$\Delta R$ for the semileptonic channels. The overall combination is less sensitive
since the $3\ell\nu + 2j$ channel is not affected by the  $\Delta R$ cut.

In several analyses \cite{JetFinding} it has been shown that significantly better
results in the identification of hadronic decays of vector boson, Higgs bosons or
supersymmetric partners of ordinary particles can be obtained analyzing the
substructure of high--$p_T$ jets in order to separate ordinary QCD jets from
those generated by the decay of heavy objects. Such an approach might overcome
the loss in discrimination between the SM and the BSM scenarios at large
$\Delta R$. We leave this possibility for further work.

\section{Conclusions}
\label{sec:conclusions}

We have examined at parton level the processes $pp \rightarrow \ell^+\ell^- + 4j$,
$pp \rightarrow 3\ell\nu + 2j$ and $pp \rightarrow \ell\nu + 4j$ including all
irreducible backgrounds contributing to these six parton final states. We have
considered three scenarios: a light Higgs SM framework with $M_H = 200 \mbox{
GeV}$, one instance of the SILH models and an infinite mass Higgs scenario in
order to determine whether the two BSM models can be distinguished from the SM
at the LHC using boson--boson scattering. For the semileptonic channels, the
largest background is $V + 4j$ at $\ordQCDsq$ which can be subtracted looking at
the distribution of the invariant mass of the two most central jets in the
region outside the weak boson mass window. We have estimated the probability, in
the two BSM scenarios, of finding, combining all set of measurements, a result
outside the 95\% probability range in the Standard Model. This probability turns
out to be about 99.9\% for the Higgsless case and 51.5\% for the SILH model.
These probabilities correspond to an integrated luminosity of $L = 200 \mbox{
fb}^{-1}$ and to the sum of all electron and muon channels for a mass of the
reconstructed pair of vector bosons larger than 600 GeV. Jet resolution plays a
crucial role in the present analysis as in all processes in which high
transverse momentum vector bosons or top particles are present and decay
hadronically.

\section *{Acknowledgments}

A.B. wishes to thank the Dep. of Theoretical Physics of Torino University
for support. \\
We gratefully acknowledge discussions with R.~Rattazzi on the
details of the SILH model.\\
This work has been supported by MIUR under contract 2006020509 004 and by the
European Community's Marie-Curie Research Training Network under contract
MRTN--CT--2006--035505 Tools and Precision Calculations for Physics Discoveries
at CollidersÕ

\vspace{2cm}



\begin{thebibliography}{999}

\bibitem{lepewwg07}
The LEP Collaborations (ALEPH, DELPHI, L3 and OPAL), 
the LEP Electroweak Working Group and the SLD Heavy Flavour Group, {\it A
combination of preliminary Electroweak measurements and constraints on the
Standard Model}, LEPEWWG/2007-01; {\tt http://lepewwg.web.cern.ch/LEPEWWG}.

\bibitem{lepewwg}
The LEP Collaborations (ALEPH, DELPHI, L3 and OPAL), 
the LEP Electroweak Working Group and the SLD Heavy Flavour Group, 
{\tt http://lepewwg.web.cern.ch/LEPEWWG}.

\bibitem{TevatronHiggsLimit}
CDF Collaboration and D0 Collaboration, 
To appear in the proceedings of 44th Rencontres de Moriond on QCD and High Energy Interactions,
La Thuile, Valle d'Aosta, Italy, 14-21 Mar 2009, arXiv:0905.2090 [hep-ex];
CDF Collaboration and D0 Collaboration,
To appear in the proceedings of 44th Rencontres de Moriond EW 2009: Electroweak Interactions
and Unified Theories, La Thuile, Italy, 7-14 Mar 2009,
arXiv:0906.1403 [hep-ex]

\bibitem{HiggsLHC} Proceedings of the Large Hadron Collider Workshop, 
Aachen 1990, CERN Report 90--10, G. Jarlskog and D. Rein (eds.). 

\bibitem{djouadi-rev1} A.~Djouadi, {\it The Anatomy of Electro--Weak Symmetry Breaking.
Tome I: The Higgs in the Standard Model}, [arXiv:hep-ph/0503172].

\bibitem{ATLAS-TDR} ATLAS Collaboration, {\it Detector and Physics Performance
Technical Design Report}, 
Vols. 1 and 2, CERN--LHCC--99--14 and CERN--LHCC--99--15.

\bibitem{Houches2003} K.A. Assamagan, M. Narain, A. Nikitenko, M. Spira, 
D. Zeppenfeld (conv.) {\it et al.}, Report of the Higgs Working Group,
Proceedings of the Les Houches Workshop on ``Physics at 
TeV Colliders'', 2003, [arXiv:hep-ph/0406152]. 

\bibitem{CMS-TDR} CMS Collaboration,{\it Technical Design Report},  
Vols. 1 and 2, CERN/LHCC 2006--001 and CERN/LHCC 2006--021.

\bibitem{reviews}
M.S.~Chanowitz, {\it Strong WW scattering at the end of the 90's:
theory and experimental prospects}.    
In {\it Zuoz 1998, Hidden symmetries and Higgs phenomena} 81-109.
[arXiv:hep-ph/9812215]

\bibitem{H_Goldstone}
  D.~B.~Kaplan and H.~Georgi,
  Phys.\ Lett.\  B {\bf 136} (1984) 183.

\bibitem{LittleH1}
  N.~Arkani-Hamed, A.~G.~Cohen and H.~Georgi,
  Phys.\ Lett.\  B {\bf 513} (2001) 232.

\bibitem{LittlestH}
N. Arkani-Hamed, A. G. Cohen, E. Katz and A.E. Nelson, JHEP 0207, 034 (2002), 
[arXiv:hep-ph/0206021].

\bibitem{gaugeHiggsU1}
  N.~S.~Manton,
  Nucl.\ Phys.\  B {\bf 158} (1979) 141;
    Y.~Hosotani,
  Annals Phys.\  {\bf 190} (1989) 233.

\bibitem{gaugeHiggsU2}
  C.~Csaki, C.~Grojean and H.~Murayama,
  Phys.\ Rev.\  D {\bf 67} (2003) 085012;
    C.~A.~Scrucca, M.~Serone and L.~Silvestrini,
  Nucl.\ Phys.\  B {\bf 669} (2003) 128.

\bibitem{HologHiggs}
  K.~Agashe, R.~Contino and A.~Pomarol,
  Nucl.\ Phys.\  B {\bf 719} (2005) 165.

\bibitem{LittleHcustodial}
S. Chang, JHEP 0312, 057 (2003), [arXiv:hep-ph/0306034].

\bibitem{Giudice_rev08}
G.F. Giudice, J.Phys.Conf.Ser.110:012014,2008, arXiv:0710.3294 [hep-ph].

\bibitem{Contino07}
  R.~Contino, T.~Kramer, M.~Son and R.~Sundrum,
  JHEP {\bf 0705} (2007) 074.

\bibitem{Giudice:2007fh}
G.F.~Giudice, C.~Grojean, A.~Pomarol, R.~Rattazzi, JHEP 0706:045,2007, 
[arXiv:hep-ph/0703164].

\bibitem{Barbieri}
R.~Barbieri, B.~Bellazzini, V.S.~Rychkov, A.~Varagnolo, Phys.Rev.D76:115008,2007,
arXiv:0706.0432 [hep-ph].

\bibitem{EEWL}
T.~Appelquist and C.W.~Bernard, \pr D22 1980 200 ; A.C.~Longhitano,
\pr D22 1980 1166 ; A.C.~Longhitano, \np  B188 1981 118;
T.~Appelquist and G.H.~Wu, \pr D48 1993 3235 (1993) [hep-ph/9304240]. 

\bibitem{history1}
M.J.~Duncan, G.L.~Kane and W.W.~Repko, \np B272 1986 517 ;
D.A.~Dicus and R.~Vega, \prl 57 1986 1110 ; 
J.F.~ Gunion, J.~ Kalinowski and A.~Tofighi--Niaki, \prl 57 1986 2351 .

\bibitem{history2}
 R.N. Cahn, S.D. Ellis, R. Kleiss and W.J. Stirling, Phys. Rev. D35 (1987) 1626;
 V. Barger, T. Han and R. Phillips, Phys. Rev. D37 (1988) 
2005 and D36 (1987) 295; R. Kleiss and J. Stirling, Phys. Lett. 200B (1988) 193;
V.\, Barger {\it et al.}, Phys.\ Rev.\ D42 (1990) 3052; 
V.\, Barger {\it et al.}, Phys.\ Rev.\, D44 (1991)  1426;
V. Barger {\it et al.},  Phys.\ Rev.\ D46 (1992) 2028; 
D.~Froideveaux, in Ref.~\cite{HiggsLHC} Vol~II, p.~444;
M.~H.~Seymour,  in Ref.~\cite{HiggsLHC} Vol~II, p.~557;
U.~Baur and E.W.N.~Glover, Phys.\ Lett.\ B252 (1990) 683;  
D. Dicus, J. Gunion and R. Vega, Phys. Lett. B258 
(1991) 475; D. Dicus, J. Gunion, L. Orr and R. Vega, Nucl. Phys. B377 (1991) 
31; J.\,Bagger {\it et al.},\pr D49 1994 1246;
V.\, Barger, R.\, Phillips and D.\, Zeppenfeld, \pl B346 1995 106 ;
J.\,Bagger {\it et al.},\pr D52 1995 3878;
K. Iordanidis and D. Zeppenfeld, \pr D57 1998 3072 ; R. Rainwater and
D. Zeppenfeld, \pr D60 1999 113004 ; erratum ibid D61 (2000) 099901.  

\bibitem{Accomando:2005hz}
  E.~Accomando, A.~Ballestrero, S.~Bolognesi, E.~Maina and C.~Mariotti,
  JHEP {\bf 0603} (2006) 093
  [arXiv:hep-ph/0512219].

\bibitem{Accomando:2006vj}
  E.~Accomando, A.~Ballestrero, A.~Belhouari and E.~Maina,
  Phys.\ Rev.\  D {\bf 75} (2007) 113006
  [arXiv:hep-ph/0603167].

\bibitem{Ambroglini:2009mg}
  G.~Bevilacqua, 
  in F.~Ambroglini {\it et al.},
  \textit{Proceedings of the Workshop on Monte
  Carlo's, Physics and Simulations at the LHC PART II}, Frascati, Italy.

\bibitem{Cheung:2008zh}
K.~Cheung, C.~Chiang and T.~Yuan,
Phys. Rev.D78(2008)051701,
arXiv:0803.2661 [hep-ph].

\bibitem{JagerOleariZeppenfeld}
B.~J\"ager, C.~Oleari and D.~Zeppenfeld, \jhep 0607 2006 015 ,[hep-ph/0603177];
B.~J\"ager, C.~Oleari and D.~Zeppenfeld,
Phys.Rev.D73:113006,2006, [hep-ph/0604200];
G.~Bozzi, B.~J\"ager, C.~Oleari and D.~Zeppenfeld,
Phys.Rev.D75:073004,2007 , [hep-ph/0701105].

\bibitem{Arnold:2008rz}
K.~Arnold {\it et al.},
arXiv:0811.4559 [hep-ph].

\bibitem{Ballestrero:2008gf}
A. Ballestrero, G. Bevilacqua, and E. Maina JHEP 05 (2009) 015, [arXiv:0812.5084].

\bibitem{Barate:2003sz}
  R.~Barate {\it et al.}  [LEP Working Group for Higgs boson searches and
  ALEPH, DELPHI, L3 and OPAL Collaborations],
  Phys.\ Lett.\  B {\bf 565} (2003) 61
  [arXiv:hep-ex/0306033].

\bibitem{Ballestrero:2007xq}
A.~Ballestrero, A.~Belhouari, G.~Bevilacqua, V.~Kashkan and
E.~Maina,
\cpc 180 2009 401 ,
arXiv:0801.3359 [hep-ph]. 

\bibitem{ref:Phase}
E.~Accomando, A.~Ballestrero, E.~Maina, 
\jhep 0507 2005 016 , [arXiv:hep-ph/0504009].

\bibitem{method} A.~Ballestrero and E.~Maina, \pl B350 1995 225 ,
[arXiv:hep-ph/9403244].

\bibitem{phact} 
A.~Ballestrero, {\tt PHACT 1.0 - \it Program for Helicity Amplitudes Calculations 
with Tau matrices'} [arXiv:hep-ph/9911318] in {\it 
Proceedings of the 14th International Workshop on High Energy Physics 
and Quantum Field Theory (QFTHEP 99)}, 
B.B.~Levchenko and V.I.~Savrin  eds. (SINP MSU Moscow), pg. 303. 

\bibitem{MadeventPaper}
F.~Maltoni, T.~Stelzer, JHEP 0302 (2003) 027;
T.~Stelzer and W.~F.~Long, Comput. Phys. Commun. {\bf 81} (1994) 357;\\
J.~Alwall {\it et al.}, JHEP 0709:028,2007, arXiv:0706.2334;\\
H. Murayama, I. Watanabe and K. Hagiwara, KEK-91-11.

\bibitem{LHAFF}
J. Alwall {\it et al.},
A Standard format for Les Houches event files.
Written within the framework of the MC4LHC-06 workshop: Monte Carlos for the LHC:
A Workshop on the Tools for LHC Event Simulation (MC4LHC), Geneva, Switzerland, 
17-16 Jul 2005,
\cpc 176 2007 300 ,
[arXiv:hep-ph/0609017].

\bibitem{CTEQ5}
CTEQ Coll.(H.L.~Lai {\it et al.}) \epj C12 2000 375 .

\bibitem{Cowan}
G.~Cowan, {\it Statistical Data Analysis}, Oxford University Press, 1998.
\bibitem{Martin:2002aw}
A.D.~Martin, R.G.~Roberts, W.J.~Stirling and R.S.~Thorne, 
Eur. Phys. J. {\bf C28} (2003) 455, [hep-ph/0211080].

\bibitem{Martin:2003sk}
A.D.~Martin, R.G.~Roberts, W.J.~Stirling and R.S.~Thorne, 
Eur. Phys. J. {\bf C35} (2004) 325, [hep-ph/0308087].

\bibitem{JetFinding}
J.M.~Butterworth,B.E.~Cox and J.R.~Forshaw, \pr D65 2002 96014 , 
[arXiv:hep-ph/0201098];\\
J.M.~Butterworth, J.~Ellis and A.R.~Raklev,
\jhep 0705 2007 033 , [arXiv:hep-ph/0702150];\\
J.M.~Butterworth,A.~Davison, M.~Rubin and G.~Salam,
Phys. Rev. Lett. 100(2008)242001,
arXiv0802.2470 [hep-ph].





\end{thebibliography}
\end{document}